\pdfoutput=1
\RequirePackage{fix-cm}
\documentclass[smallextended]{svjour3}       % onecolumn (second format)
\smartqed  % flush right qed marks, e.g. at end of proof
\usepackage{graphicx}
 \usepackage[utf8]{inputenc}
\usepackage{amssymb}
\usepackage{amsmath}
\usepackage{esvect}
\usepackage{csquotes}
\usepackage[dvipsnames]{xcolor}
\usepackage{kmath}
\usepackage{float}
\usepackage{hyperref}
\renewcommand{\PACSname}{\textbf{JEL classification}\enspace}
\usepackage{color}

\DeclareMathOperator*{\argmax}{arg\,max}

\begin{document}

\title{Decision-facilitating information in hidden-action setups: An agent-based approach%\thanks{This work is funded by the Anniversary Fund of the Oesterreichische Nationalbank under project No. 17930.}
}


\author{Stephan Leitner \and Friederike Wall
}


\institute{
Stephan Leitner\at
              University of Klagenfurt\\
              Department of Management Control and Strategic Management\\
              Universit\"atsstra{\ss}e 65-67\\
              9020 Klagenfurt, Austria \\
              \email{stephan.leitner@aau.at}           \\
              ORCID: 0000-0001-6790-465
           \and
      	Friederike Wall\at
              University of Klagenfurt\\
              Department of Management Control and Strategic Management\\
              Universit\"atsstra{\ss}e 65-67\\
              9020 Klagenfurt, Austria \\
              \email{friederike.wall@aau.at} \\
              ORCID: 0000-0001-8001-8558      
}

\date{Received: date / Accepted: date}

\maketitle

\begin{abstract}
The hidden-action model captures a fundamental problem of principal-agent theory and provides an optimal sharing rule when only the outcome but not the effort can be observed \cite{holmstrom1979}. However, the hidden-action model builds on various explicit and also implicit assumptions about the information of the contracting parties. This paper relaxes key assumptions regarding the availability of information included in the hidden-action model in order to study whether and, if so, how fast the optimal sharing rule is achieved and how this is affected by the various types of information employed in the principal-agent relation. Our analysis particularly focuses on information about the environment and \textcolor{black}{about} feasible actions for the agent. We follow an approach to transfer closed-form mathematical models into agent-based computational models and show that the extent of information about feasible options to carry out a task only has an impact on performance if decision makers are well informed about the environment, and that the decision whether to perform exploration or exploitation when searching for new feasible options only affects performance in specific situations. Having good information about the environment, on the contrary, appears to be crucial in almost all situations. 
\keywords{Management control \and complexity economics \and agent-based simulation \and information systems \and information system sophistication \and search strategy}
 \PACS{C63 \and D83 \and D86}
\end{abstract}

\section{Introduction}

\textcolor{black}{
This paper focuses on the hidden-action model introduced in \cite{holmstrom1979} (henceforth referred to as standard hidden-action model): An individual (the principal) delegates some authority in order to act in her name (i.e., to exert productive effort) to another individual (the agent). This relation is specified in a contract which defines what the agent has to do, in terms of a task which is delegated from the principal to the agent, and how the resulting outcome is shared between the principal and the agent \cite{Lambert2001,eisenhardt1989}. The key issue in the standard hidden-action model is that the agent's productive effort cannot be observed by the principal. In order to cope with this problem, a sharing rule is proposed which is based on outcome which might be measured by a particular performance metric.  As a consequence of the proposed sharing rule, research related to this model often puts particular emphasis on information systems which the principal employs for measuring the outcome which was achieved by the agent fulfilling the task (for an overview see, for example, \cite{Baiman1982,Baiman1990,Hesford2007,Lambert2001,Lambert2006}).} 

\textcolor{black}{
While the standard hidden-action model puts strong emphasis on information to control the behavior of the agent through decision-influencing information, the role of information to reduce pre-decision uncertainty in order to increase the probability of better decisions by means of decision-influencing information remains widely unconsidered \cite{Demski1976}: By giving the principal and the agent all the information required to make optimal decisions and by making rather \enquote*{heroic} assumptions related to the principal's and the agent's information-processing capabilities \cite{mueller1995,axtell2007,simon1959,simon1979}, most of the problems related to the availability of decision-facilitating information are \enquote*{assumed away}.} For example, the principal as well as the agent are assumed to have knowledge about the distribution of the states of the relevant environment which, together with the productive effort exerted by the agent, shapes performance. Moreover, the principal is assumed to be perfectly informed about the agent's utility function and to have knowledge about the entire set of feasible actions from which the agent selects the level of productive effort which he exerts, which also means that the principal is able to \textit{immediately} find the optimal sharing rule. There is, however, empirical evidence that the assumptions incorporated in the standard hidden-action model are not perfectly in line with human capabilities and human behavior, particularly when studied in the context of organizations (e.g., \cite{perrow1986,eisenhardt1989,hendry2002}). 

In this paper, we shift the attention from the decision-influencing perspective to the decision-facilitating role of information. In particular, we focus on the assumptions related to decision-facilitating information and analyze whether a relaxation of the aforementioned assumptions results in modified requirements for the information which is relevant in the context of the standard hidden-action model. In particular, employing more realistic assumptions regarding the availability of decision-facilitating information could shift the attention from ex-post performance information to information systems which provide information about the environment or the set of feasible actions. Limiting the principal's and agent's availability of decision-facilitating information will inevitably require to endow them with additional capabilities, such as learning mechanisms or the ability to search for information following different strategies, such as exploration or exploitation \cite{march1991}. 

There are some empirical findings supporting the conjecture that managerial decision makers have particular requirements regarding \textcolor{black}{decision-facilitating} information: For example, Vandenbosch and Huff \cite{Vandenbosch1997} report on the use of executive information systems indicating that managers employ the systems to challenge general managerial assumptions and preconditions, e.g., related to the environment. Based on an empirical study, Sch\"affer and Steiners \cite{Schaeffer2004} distinguish different forms of information usage indicating that ex post performance evaluation is just one of several relevant forms while, according to \cite{Schaeffer2005}, managers are rather satisfied with accounting-based information but see some deficiencies with respect to information related to environment including, for example, competitors or probabilities of external events. In a similar vein, Hall \cite{Hall2010} argues that future developments in accounting should be directed to provide relevant information for a general understanding of the related field and for strategizing.

Against this background, we employ an approach which allows for a relaxation \cite{guerrero2011,Leitner2015a} of key assumptions regarding the availability of decision-facilitating information of the contracting parties in the standard hidden-action model: We put a particular focus on the assumptions related to (i) the availability of information about the environment and the assumption that (ii) both the principal and the agent  are fully informed about the set of actions feasible to carry out the task which the principal delegates to the agent. 
%
%\textcolor{black}{We relax the assumptions included in the standard hidden-action model in the following way:}
%\begin{itemize}
%\item We \textcolor{black}{limit the availability of information and} endow the principal and the agent to learn about (i) the environment over time: For the agent we carry over the assumption of the \textcolor{black}{standard hidden-action model} and endow him with the capability to observe realizations of the environments ex-post to their realization, while the principal is endowed with the ability to ex-post estimate the environmental variables. Both, the principal and the agent have access to private information systems in which their learnings are stored. 
%\item For (ii) the set of feasible actions we add information asymmetry: The agent is fully informed about feasible ways to carry out the delegated task. The principal, however, only has limited information about the set of feasible actions but is endowed the ability to either perform exploration or exploitation to overcome this information asymmetry. Both, the principal and the agent have private information systems from which information about the action space can be retrieved. 
%\end{itemize}
The relaxation of assumptions reduces the model's mathematical tractability dramatically. We therefore set up an \textcolor{black}{agent-based variant} of the hidden-action problem which includes the relaxed assumptions (for simulation-based approaches in managerial science see, for example, \cite{Davis2007,Leitner2015,Wall2016}).

The remainder of this paper is organized as follows: Section \ref{sec:2} introduces the main features of the hidden-action model introduced in \cite{holmstrom1979} and discusses the assumptions incorporated into the standard hidden-action model. In Section \ref{sec:3}, we introduce the relaxed assumptions, discuss their operationalization, and formalize the agent-based model. Section \ref{sec:4} presents and discusses the results of the simulation study. Section \ref{sec:5} summarizes and concludes the paper, discusses limitations and avenues for future research on this topic. 

\section{Information and information systems in the standard hidden-action model}
\label{sec:2}

\subsection{The hidden-action model in a nutshell}
\label{sec:ha-nutshell}

The standard hidden-action model \textcolor{black}{is a static model} that captures a single-period situation in which a principal hires an agent for carrying out a task.\footnote{Please note that we focus on the model variant introduced in \cite{holmstrom1979}. Other model versions are, for example, introduced in \cite{Harris1979} and \cite{Spence1971}.}   \textcolor{black}{The principal's main role is to supply capital and to construct incentives, while the agent's main role is to act on behalf of the principal. The hidden-action model can be applied to a multiplicity of situations, such as economic relationships between employer and employee, homeowner and contractor, or buyer and supplier \cite{Caillaud2000,eisenhardt1989}. The concept of the task delegated from the principal to the agent is, therefore, rather abstract and universal.}

\textcolor{black}{\paragraph{Main characteristics of the standard hidden-action model.} The main features that describe the economic relationship between the principal and the agent are the following \cite{holmstrom1979,Caillaud2000}:
\begin{itemize}
\item The principal delegates a task to the agent. 
\item The agent exerts effort (also referred to as action) to carry out the task, which affects the principal's payoff. We denote the set of all feasible effort levels by \(a\in\mathbf{A} \subseteq \mathbb{R}\). The agent has full power to decide for an effort level out of a set of feasible effort levels \(\mathbf{A}\). 
\item The agent's effort \(a\) is hidden. This means that only the agent knows about the effort exerted; the principal cannot observe it (at least not at reasonable costs). Information about the agent's effort is, thus, distributed asymmetrically between the principal and the agent.
\item The effort \(a\) together with a random state of nature \(\theta\) determines the task's outcome \(x=x(a,\theta)\).  The outcome \(x\) can be observed by both the principal and the agent. The state of nature \(\theta\) is a random variable that describes the economic environment in which the task is carried out. It cannot be observed by the principal. The agent, however, can observe (or deduce) the state of nature after the outcome \(x\) has taken effect. 
\item Due to a lack of observability of the effort \(a\) to the principal, incentives for the agent have to be functions of the outcome \(x\) alone. The principal and the agent agree \textit{ex ante} on a rule (an incentive scheme) that defines how the task's outcome \(x\) is shared between the two parties. We denote the sharing rule by \(s(\cdot)\) and the agent's share of outcome \(x\) by \(s(x)=x\cdot p\), whereby \(p\) stands for the premium-parameter.
\item The principal is assumed to be risk-neutral. Her utility is defined over payoff and the agent's compensation. We denote her utility function by \(U_P(x-s(x))=x-s(x)\). 
\item The agent is assumed to be risk-averse. His utility is defined over compensation and disutility for the effort exerted. We denote his utility function by \(U_A(s(x),a)=V(s(x))-G(a)\), with \(V^\prime>0\) and \(x_a \geq 0\).\footnote{Subscript \(a\) denotes the partial derivative with respect to \(a\)} 
\end{itemize}}

\noindent Figure 1 represents the sequence of events within the standard hidden-action model.  \textcolor{black}{In \(\tau=1\), the principal designs a contract and offers it to the agent, who in \(\tau=2\) decides whether to accept the contract or not. The contract specifies the task and the sharing rule, and is valid for one period. Once the contract is accepted it is binding for both parties. } If the agent accepts the contract, he exerts \textcolor{black}{(and completes)} productive effort in \(\tau=3\). \textcolor{black}{The model is strictly sequential, which prevents the agent from adjusting his effort after \(\tau=3\).} The random state of nature \(\theta\) realizes  in \(\tau=4\). Finally, in \(\tau=5\), the outcome and the principal's and the agent's utilities take effect. 

\begin{figure}
 \center \includegraphics[width=0.7\linewidth]{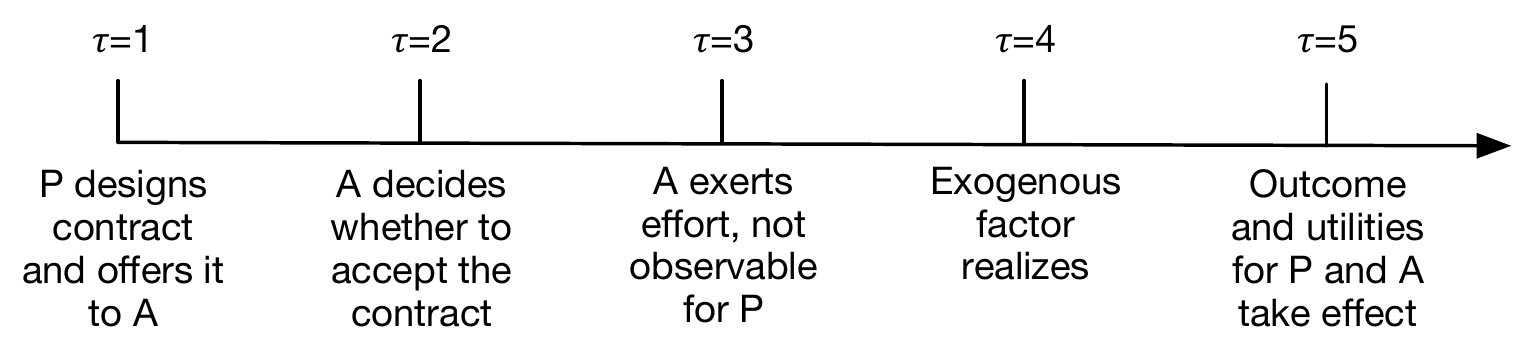}
  \caption{Sequence of events within the standard hidden-action model (P=principal, A=agent)}
  \label{fig:1}
\end{figure}

\paragraph{\textcolor{black}{Constraints to be considered from the principal's point of view.}} The principal's decision problem is to find a sharing rule \(s(\cdot)\)
\begin{itemize}
\item that meets a minimum level of utility \(\underline{U}\) for the agent. \textcolor{black}{\(\underline{U}\) is referred to as reservation utility and represents the expected utility from pursuing his next best (outside) alternative. This requirement for the sharing rule assures that the agent accepts the contract in \(\tau=2\). In the hidden-action model, this condition is referred to as participation constraint \cite{holmstrom1979,Caillaud2000}.} 
\item \textcolor{black}{that motivates the agent to exert a level of effort that maximizes both the principal's and the agent's utilities. In the design of the sharing rule, the principal has to take into account that the rule affects the agent's choice of effort \(a\) via outcome \(x\), and that effort \(a\) leads to disutility for the agent. This condition is referred to as incentive compatibility constraint \cite{holmstrom1979,Caillaud2000}.}
 \end{itemize}

\textcolor{black}{\paragraph{The optimization program.} Given the main characteristics of the standard hidden-action model and the constraints to be considered as described above, the optimization program to generate pareto-optimal sharing rules can be formalized by
\begin{subequations}
\begin{align}
\max_{s(x),a} 	\quad 	& 	{E}\left(U_P\left(x-s\left(x\right)\right)\right) \label{eq1:maximization}\\
\textrm{s.t.} 	\quad 	&	{E}\left(U_A\left(s\left(x\right),a\right)\right)\geq\underline{U} \label{eq1:PC}\\
 					 &	a \in \argmax_{a^\prime \in \mathbf{A}}  {E} \left(U_A\left(s\left(x\right),a^\prime\right)\right)~, \label{eq1:ICC}
\end{align}
\end{subequations}
\noindent where Eqs. (\ref{eq1:PC}) and (\ref{eq1:ICC}) represent the participation and the incentive compatibility constraints, respectively. The notation \enquote*{\(\argmax\)} denotes the set of arguments that maximize the objective function that follows, \({E}\) denotes the expectation operator \cite{holmstrom1979}. The solution to the standard hidden-action model, as formalized in Eqs. (\ref{eq1:maximization})-(\ref{eq1:ICC}), is presented in Appendix \ref{app:c}. Table \ref{tab:notation-standard} summarizes the notation used in this section.}

\begin{figure}
  \includegraphics[width=\linewidth]{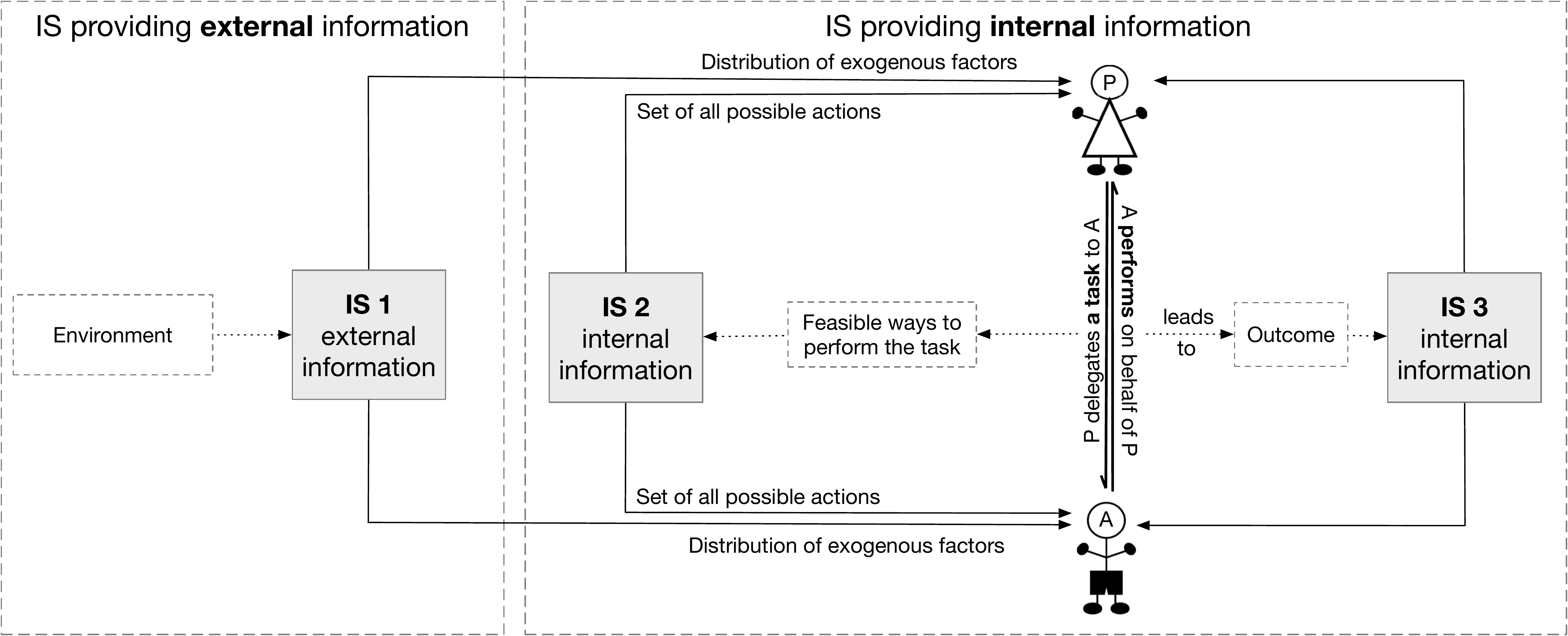}
  \caption{Information systems in the standard hidden-action model (P=principal, A=agent)}
  \label{fig:2}
\end{figure}

 \begin{table}
 \textcolor{black}{
  \caption{Notation used in the (static) standard hidden-action model}
     \label{tab:notation-standard}
 \begin{tabular}{p{0.57\textwidth}p{0.35\textwidth}}
\hline\noalign{\smallskip}				
Key element & Notation \\
\hline\noalign{\smallskip}
Principal's utility function						&	\(U_P(x-s(x))\)				\\
Agent's utility function						&	\(U_A(s(x),a)\)				\\
Agent's utility from compensation				&	\(V(s(x))\)					\\
Agent's disutility from effort					&	\(G(a)\)					\\
Agent's reservation utility						&	\(\underline{U}\)				\\
Agent's share of outcome						&	\(s(x)=x\cdot p\) 			\\
Outcome									&	\(x=x(a,\theta)\)				\\
Premium parameter							&	\(p\)						\\
Effort level								&	\(a\)						\\
Set of feasible effort levels					&	\(\mathbf{A}\)				\\
Random state of nature						&	\(\theta\)					\\
\hline\noalign{\smallskip} 
\end{tabular}
}
\end{table}

\subsection{Concepts of information and information asymmetry}
\label{ha:info}

\textcolor{black}{The concept of information is central to the hidden-action model. In contrast to the competitive general equilibrium view that assumes that information is given and perfectly known (e.g., \cite{Arrow1964}), the principal-agent theory follows the stream of information economics and regards information as being imperfect and costly \cite{Stiglitz2003,Frieden2010,Hawkins2010}. Consequently, in the latter stream of research asymmetries of information play a central role.}

\textcolor{black}{The concept of information has a long history, is often domain-dependent, and is applied in multiple ways in the research literature (for an overview see, for example, \cite{Capurro2009,Hofkirchner2009,madden2000,McCreadie1999}). At a very general level, information can, for example, be defined as a fact which one is told, systematic data that conveys a message, or a numerical quantity that measures the uncertainty in the outcome of an experiment \cite{madden2000,Khan2018,Soofi1994}. Notions of information capture a wide spectrum ranging from semantic to technical definitions: The former notions use information in a rather intuitive sense while in the latter ones information is usually a well-defined function to capture the extent of uncertainty \cite{Soofi1994}. McCreadie and Rice \cite{McCreadie1999} rather focus on semantic notions and provide a more detailed overview of conceptualizations and distinguish between information as 
\begin{enumerate}
\item \textbf{a resource or commodity}: McCreadie and Rice \cite{McCreadie1999} argue that information can be produced, purchased, replicated, distributed, manipulated, passed or not passed along, controlled, traded, and sold. This concept of information is consistent with the very well-known sender-receiver model of communication \cite{Shannon1948,Shannon2001} and, therefore, might also include assumptions about the interpretation and understanding of information by the receiver. It is, however, also recognized that information is different from other commodities, as it possesses properties of public goods: Its consumption is, for example, non-rivalrous and from the macro-perspective it is not efficient to exclude it from others \cite{Stiglitz2000}. According to \cite{McCreadie1999}, this conceptualization might be used as a basis for strategies for generating and consuming information (e.g., in the context of innovation and problem-solving) and for establishing the value of information. Even though it is inefficient from a macro-perspective, this conceptualization constitutes the basis for \textit{individual} benefits resulting from the availability of private information. 
\item \textbf{data in the environment}: According to this concept, information is available in the environment (e.g., in terms of smells, sounds, artifacts, objects) and interacts with human information-processing capabilities. This view on information includes unintentional communication that is driven by perception \cite{Buckland1990} and is also related to the understanding that the value of information will be judged on the basis of the environment in which it is used: relevant environments can, e.g., be characterized alongside the dimensions of people (e.g., roles, cognitive abilities), problems (e.g., degree of structuring, experienced problem dimensions), settings (e.g., organization, means of communication), and problem resolutions (e.g., the way information is used, rules for solving problems) \cite{Taylor1991,katzer1992,Khan2018}.
\item \textbf{representation of knowledge}: This concept of information views information as a representation of (or is a pointer to) knowledge \cite{McCreadie1999}. A recent example of this concept of information is persistent digital identifiers for scholarly publications, such as ORCID or DOI. 
\item \textbf{part of the process of communication}: Here, information is seen as part of human behavior, and meanings are rather embodied in people than in words \cite{madden2000}. This concept shifts the focus from data to the way users handle data \cite{Budd1987} and also includes temporal, social, and personal factors as parts of work practices into the process of information acquisition and communication \cite{McCreadie1999}. This view on information is, for example, reflected by the well-known distinction between implicit and explicit knowledge \cite{Dienes1999}.
\end{enumerate}}

%\textcolor{black}{\paragraph{Roles of information.} Demski and Feltham \cite{Demski1976} distinguish between two roles of (accounting) information, namely the decision-influencing and the decision-facilitating role. 
%%
%\textit{Decision-facilitating information} is an input into the decision-making process and is expected to help to make better decisions \cite{Evans1994,Leitner2012}. Better decisions are, for example, achieved by reducing ex-ante uncertainty \cite{Sprinkle2003}, by a revision of decision makers' beliefs based on new information \cite{Baiman1982}, or by helping assisting in problem solving \cite{Emsley2005}. 
%%
%\textit{Decision-influencing information}, on the contrary, is used ex-post to the decision. This type of information is, for example, used to overcome problems of behavioral control that arise from selfish behavior \cite{Baiman1982}.  
%The hidden-action model has a very strong focus on the decision-influencing role of information: it is mostly the agent's exerted effort \(a\) that is in the center of interest. Problems of decision facilitating information are assumed away by giving the principal and the agent full access to all relevant information that is used to make decisions, such as information about the environment, the individual utility functions and the production function.} 

\textcolor{black}{\paragraph{Asymmetric information in the hidden-action setup.} The standard hidden-action model uses the concept of information as good or commodity \cite{McCreadie1999} and particularly focuses on the decision-influencing role of information \cite{Demski1976}. The principal is unable to observe or verify at reasonable costs the effort \(a\) made by the agent. The agent, of course, knows the effort level \(a\) \cite{Spremann1987}. In addition, only the agent but not the principal can observe the exogenous variable \(\theta\) that realizes in \(\tau=4\) (see Fig. \ref{fig:1}). The latter assumption assures that there is, indeed, asymmetry in information about effort \(a\), as the principal \textit{cannot} perfectly infer \(a\) from outcome \(x\) without knowing \(\theta\) \cite{Caillaud2000}. Please note that with respect to all other elements of the standard hidden-action model introduced in Sec. \ref{sec:ha-nutshell}, the principal and the agent share the same state of information, i.e., there is only information asymmetry with respect to the effort \(a\) and the realized exogenous factor \(\theta\). The fact that (i) the outcome \(x\) is correlated to \(a\) and (ii) that it is observable for both the principal and the agent without costs opens up a way to overcome the asymmetry in information for the principal: She designs the reward scheme \(s(\cdot)\), which provides incentives for the agent to make a level of effort that does not only maximize his own but also the principal's utility. This notion of information asymmetry is used in both the static hidden-action model (see Sec. \ref{sec:ha-nutshell}) and the dynamic agent-based representation of the hidden-action model (see Sec. \ref{sec:3}).   }

\textcolor{black}{\paragraph{Information in the agent-based model.} The agent-based model introduced in this paper also follows the concept of information as good or commodity. Recall that the standard hidden-action model assumes information other than the effort level \(a\) and the exogenous variable \(\theta\) to be given and available for both the principal and the agent; the transfer to the agent-based model allows for limiting the availability of these pieces of information and, therefore, for shifting the focus from the decision-influencing to the decision-facilitating role of information. At the same time, the dynamic features of the agent-based model allow for endowing the principal and the agent with learning capabilities, so that they can assemble the missing pieces of information over time. Thus, we extend the notion of asymmetric information (as introduced above) by the difference between information \(J\), which is intrinsic to a system and information \(I\), which represents information about a system \cite{Frieden2010,Hawkins2010}: \(J\) stands for the most complete and perfectly knowledgeable information \textit{intrinsic to a system} while \(I\) can be interpreted as information \textit{about a system} based on a number of observations. In the agent-based model variant we particularly consider that information about the environment and the feasible actions to carry out the delegated task are not given but are to be learned or discovered by the principal and the agent over time. Information \(J\) could, for example, represent the distribution of exogenous variables and \(I\) could stand for the realizations of the exogenous variable observed by the agent. An observation can, therefore, be modeled as an information-flow process \(J\rightarrow I\).\footnote{Please note that we focus on the concept of information as resource or commodity and, therefore, exclude effects of cognitive abilities at the agent-level, which would alter the information-flow process.} Increasing (decreasing) the number of observations decreases (increases) the distance \(J-I\) and therefore allows for agents being better (less) informed about \(J\), which reflects some of the results presented in \cite{Kreps1979,Puppe1996,Billot1999}.\footnote{Please note that the notion of asymmetric information used in the standard hidden-action model is a binary modeling choice, i.e., information about the effort level \(a\) is either available or not available. The extended notion of information introduced here allows for being better (or less) informed (cf. also information entropy \cite{Shannon1948}). Please also note that if the number of observations gets sufficiently large there is a chance that additional observations do not contain any additional information \cite{Billot1999}. } 
Both the principal and the agent store their observations in information systems. The different types of information systems (implicitly) included in the standard hidden-action model are discussed in the next section.   }

\subsection{Information systems in the hidden-action model}
\label{ha:is}

The standard hidden-action model in total captures (in parts implicitly) three types of information systems (ISs), \textcolor{black}{which provide the principal and the agent with relevant pieces of information to make optimal decisions} (cf. Fig. \ref{fig:2}). We distinguish between internal and external information. Internal information \textcolor{black}{is related to the business of an organization} and captures, for example, information \textcolor{black}{coming form an organization's planning systems} or information about an organization's performance. External information refers to information about the \textcolor{black}{economic environment in which the task is carried out.} The following ISs are considered: 
\begin{enumerate}
\item \textbf{Information about the environment (provided by IS 1)}: The principal and the agent are modeled to be informed about the distribution of the exogenous factor \(\bold{\theta}\). This type of information covers all relevant issues outside the organization, such as information about competitors, resources, technology, economic conditions \cite{dumond1994}: The principal uses this information in \(\tau=1\), when \textcolor{black}{she fixes the incentive scheme} \(s(\cdot)\), while the agent uses this information in \(\tau=2\) and \(\tau=3\), when he decides whether to accept the contract or not and selects an effort level \(a\), respectively.
\item \textbf{Information about the action space (provided by IS 2)}: In order to derive the optimal sharing rule \textcolor{black}{\(s(\cdot)\)} \textcolor{black}{instantaneously} in \(\tau=1\), and in order to exert the utility maximizing effort level in \(\tau=3\), both the principal and the agent have to have information about the entire action space \(\bold{A}\) \textcolor{black}{which is provided by IS 2.} In order to find the optimal compensation scheme \(s(\cdot)\) and to find the optimal effort level \(a\), the principal and the agent are required to consider all relevant alternatives within the action space. \textcolor{black}{This assumption is rather \enquote*{heroic}} as the specification of \textit{all} {feasible} effort levels might be extremely difficult \cite{feltham1968}.
\item \textbf{Information about the outcome (provided by IS 3)}: In \(\tau=5\) the standard hidden-action model assumes the principal and the agent to observe outcome \(x\). In order to do so, they use an IS which also provides {internal} information. 
\end{enumerate}

Figure \ref{fig:2} schematically represents ISs \textcolor{black}{included} in the hidden-action context. 
This paper  focuses on the decision-facilitating role of information in the context of hidden-action problems. For the remainder of the paper, we particularly stress the assumptions regarding the ISs which provide external information (IS 1) and internal information about the action space (IS 2). We do not focus on IS 3, as the type of information provided by this information system is mainly used for decision-influencing purposes. It is also important to note that the standard hidden-action model assumes that the different types of ISs contain \textit{all} the necessary pieces of information. For the \textcolor{black}{standard hidden-action} model described above, this means that \textit{all} required information is entered into ISs 1 and 2 before \(\tau=1\). 

\section{The agent-based model variant}
\label{sec:3}

\subsection{Relaxing assumptions regarding information systems}
\label{sec:relaxation}

\subsubsection{Assumptions regarding the IS which provides external information (IS 1)}
 \label{ha:is1}

\begin{figure}
  \includegraphics[width=\linewidth]{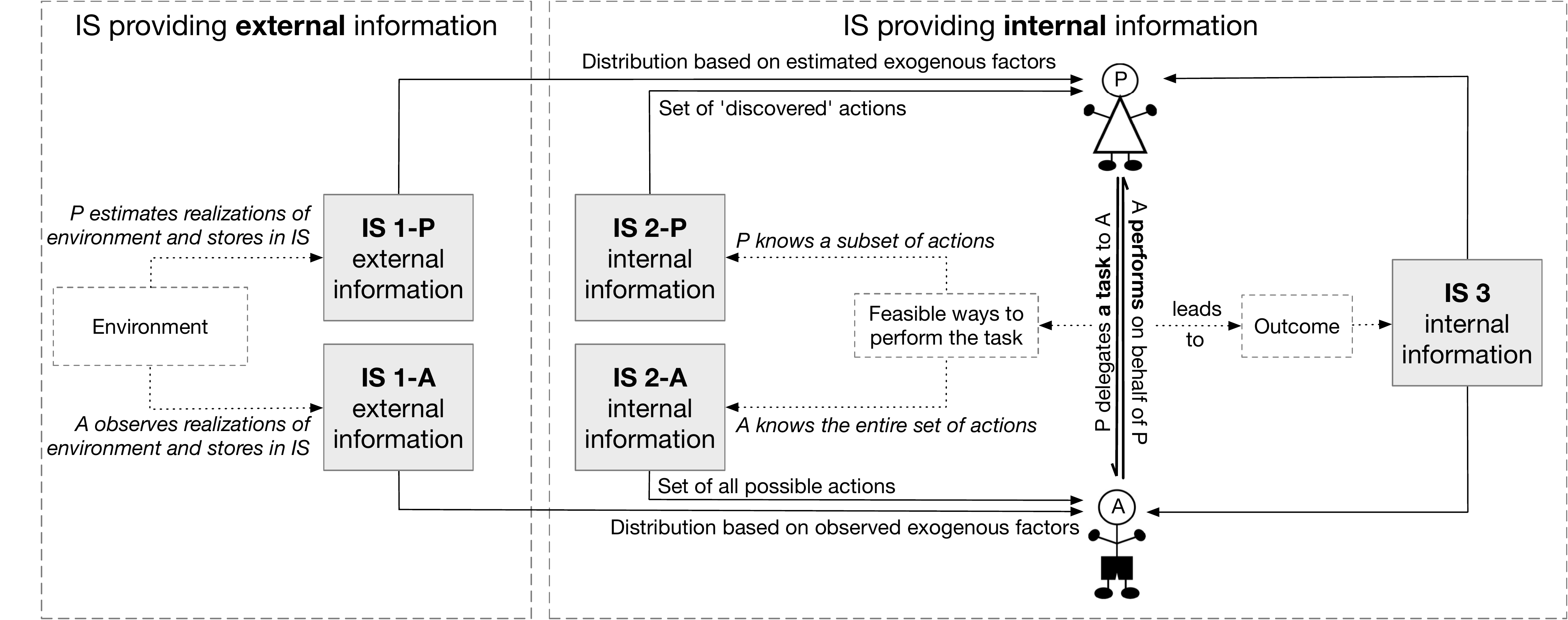}
  \caption{Information systems in the agent-based model variant (P=principal, A=agent)}
  \label{fig:3}
\end{figure}

\paragraph{Relaxed Assumptions.} Recall that the standard hidden-action model (introduced in \cite{holmstrom1979}) assumes that both the principal and the agent \textit{instantaneously} have {information about the distribution of the exogenous factor capturing the environment. From a decision-facilitating perspective this means that the principal and the agent can immediately make the best possible decision.} We adapt this assumption in the following way: 
\begin{enumerate}
\item  The principal and the agent no longer instantaneously have information about all possible realizations of environmental variables as assumed \textcolor{black}{by} the standard hidden-action model.
\item  The principal and the agent are endowed with the \textit{individual} capability to learn about the environment (i.e., about the distribution of environmental variables) \textit{over time}. We refer to the implemented learning model as simultaneous sequential learning. 
\item  The principal and the agent \textcolor{black}{no longer share one IS but} can store their individually acquired pieces of information IS 1-P and IS 1-A, respectively (cf. Fig. \ref{fig:3}).  
\end{enumerate}
These are feasible adaptations of the standard model, as learning about the environment is a common feature of organizations: Epstein \cite{epstein2003}, for example, refers to it as a prerequisite for organizational well-being and survival (see also \cite{draft1984,guo2018}). 

\paragraph{Operationalization of relaxed assumptions.} In order to operationalize these adaptations, we enrich the standard hidden-action model \textcolor{black}{by} a simultaneous and sequential learning model: We endow the principal with the ability to estimate the realizations of \(\theta\), which she stores in her private IS 1-P. The principal's learning mechanism is formalized in Sec. \ref{sec:abm}. For the agent we transfer the assumption of the standard hidden-action model, so that he is able to (\textcolor{black}{ex-post, in \(\tau=5\)}) observe the realizations of \(\theta\), which he stores in his private IS 1-A. For a schematic representation of the \textcolor{black}{related ISs included in the agent-based model variant} cf. Fig. \ref{fig:3}. 

\paragraph{Sophistication of IS 1-P and IS 1-A in the agent-based model variant.} Our model considers different sophistication levels of the ISs which provide external information. The concept of sophistication of IS 1-P and IS 1-A can be related to prior research in two ways: 
\begin{enumerate}
\item Guo and Reithel \cite{guo2018} divide information-processing in organizations into two main categories, namely information inflow and information outflow. We capture the inflow of information about the environment into the IS by the principal and the agent entering their learnings (\textcolor{black}{according to} the simultaneous sequential learning model) into \textcolor{black}{their ISs}. As a prerequisite for information outflow, the collected information, \textcolor{black}{which was previously entered into the ISs,} needs to be processed so that it can be used for decision-making purposes. IS sophistication is referred to as the organization's capability to process information.
\item The concept of IS sophistication can also be related to the fit between the individual, the task to be carried out, and the IS providing information: Liu et al. \cite{liu2011} break down the task-technology fit model (\textcolor{black}{which} is originally introduced in \cite{goodhue1995}) and argue that there are  three two-way fits, namely the task-technology fit, the individual-technology fit, and the task-individual fit. The latter fit refers to the fit between individual capabilities and decision-making requirements imposed by certain tasks. The individual-technology fit refers to the fit between characteristics of technologies and needs of individuals who are responsible for solving tasks, e.g. in terms of how information is provided. The task-technology fit refers to the fit between characteristics of technologies and the tasks to be carried out, for example in terms of providing good and appropriate information. \textcolor{black}{Information system sophistication  is conceptually related to} the task-technology fit.
\end{enumerate}

We operationalize the sophistication of the ISs which provide external information as follows: More (less) sophisticated ISs comprise better (poorer) information about the environment, whereby the lowest (highest) level of sophistication indicates that only very recent (all historical) data stored in the IS is processed and provided for decision-making purposes. \textcolor{black}{Recall the concept of information used in the agent-based model introduced in Sec. \ref{ha:info}: If an IS is characterized by a high level of sophistication, this means that it can provide more reliable information, in terms of information \(I\), about the information \(J\) that is intrinsic to a system. A higher (lower) level of sophistication, thus, decreases (increases) the distance \(J-I\).}
 
 \subsubsection{Assumptions regarding the IS which provides internal information (IS 2)}
 \label{ha:is2}

\paragraph{Relaxed assumptions.} The assumptions of the standard \textcolor{black}{hidden-action} model imply that there is one IS which comprises information about the set of feasible actions \(\bold{A}\), and that both the principal and the agent have the same information about the action space \textcolor{black}{available} (cf. Fig. \ref{fig:2}). Prior research, however, argues that information asymmetry \textcolor{black}{(following the concept of information asymmetry introduced in Sec. \ref{ha:info})} lies in the heart of decentralization \cite{akerlof1970}, as decentralized decision makers (i.e, the agent) are usually better informed than central managers or the owner of an organization (i.e., the principal) \cite{rajan2006}. In line with the argumentation in \cite{akerlof1970} we enrich the standard hidden-action model with information asymmetry about the set of feasible actions between the principal and the agent, so that we allow for the agent to be better informed about the action space than the principal.

\paragraph{Operationalization of relaxed assumptions.} In order to introduce information asymmetry regarding \(\bold{A}\), we make the following adaptations of the standard hidden-action model: 
\begin{enumerate}
\item The principal and the agent no longer share one IS which contains information about the set of feasible actions but we model \textcolor{black}{them} to have \textit{separate} ISs (IS 2-P and IS 2-A in Fig. \ref{fig:3}). The fact that the principal and the agent no longer share the same information allows for the agent to be better informed about \textcolor{black}{the set of feasible actions} \(\bold{A}\) than the principal. 
\item As the principal no longer has full information about the set of feasible actions, we endow her with the ability to either search locally (exploitation of the known action space) or globally (exploration outside of the known action space) for actions which she wants the agent to carry out \cite{march1991}. 
\end{enumerate}

\paragraph{Sophistication of IS 2-P and IS 2-A in the agent-based model variant.} Our operationalization of information asymmetry is in line with previous research: Rajan and Saouma \cite{rajan2006}, for example, argue that the extent of information asymmetry is influenced by the choice of the internal accounting system. We take up this argumentation and set up the model in the following way: While the agent has access to the entire action space \(\mathbf{A}\) using his private IS 2-A, we model the principal to be able to see only a fraction of \(\mathbf{A}\) as a consequence of the sophistication of her private IS 2-P. A low (high) level of sophistication of IS 2-P, thus, indicates that the principal has information about a relatively small (large) fraction of \(\bold{A}\).\footnote{Please be aware that our adaptation blurs the line between two types of principal agent models: As outlined above, in the standard hidden-action model all information except the effort \(a\) made by the agent is available for both the principal and the agent. In the hidden-information scenario, the principal and the agent share all information except some observations which only the agent has made. Arrow \cite{arrow1985} argues that the observations (which lead to private information for the agent), for example, relate to possibilities of production which are not available for the principal. This argumentation can be directly related to our operationalization of the sophistication of the IS which provides information about the set of feasible actions: An increase (decrease) in the amount of the agent's private information (as a consequence of observations) leads to the agent being better informed about the set of feasible actions \(\bold{A}\), which can be translated into a low (high) level of sophistication of the internal information system.} 

\subsection{Formalization of the agent-based model variant}
\label{sec:abm}

\begin{figure}
\center
  \includegraphics[angle=90,scale=0.45]{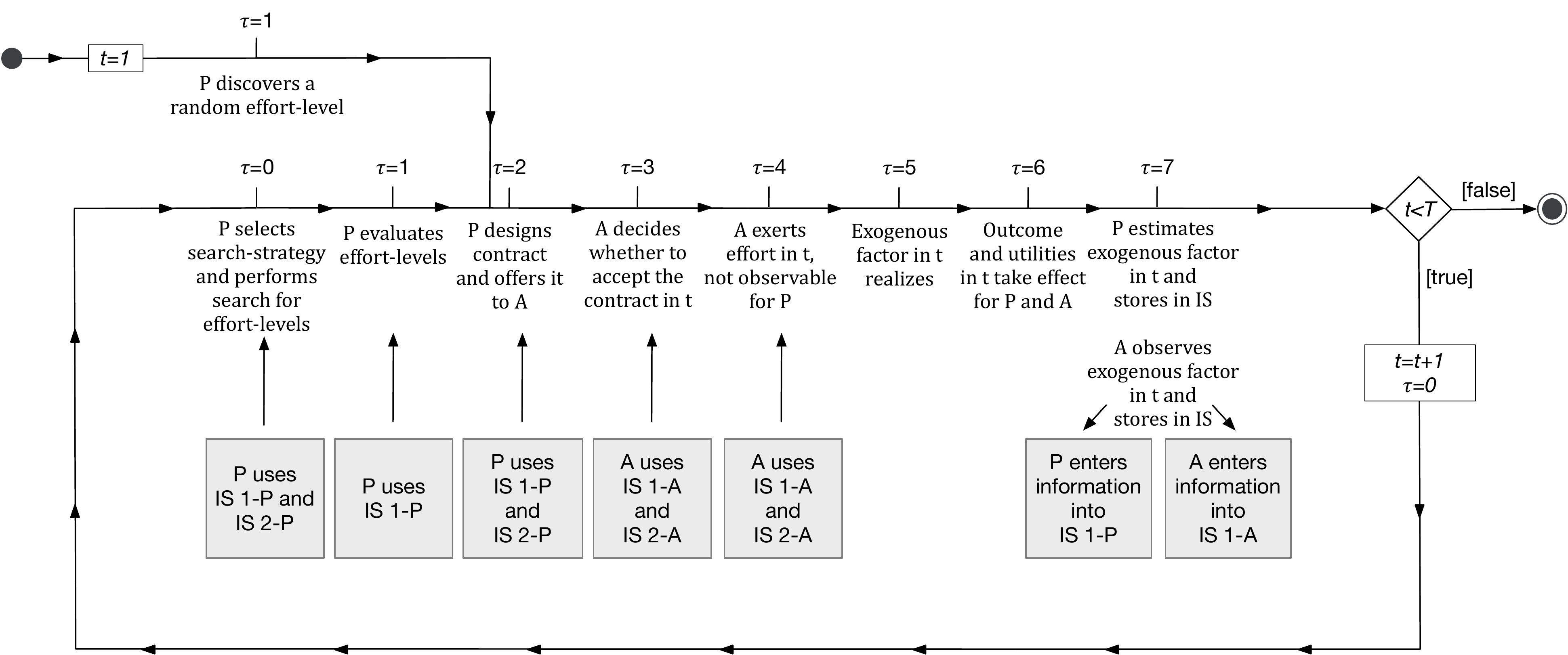}
  \caption{Sequence of events \(\tau\) for one timestep \(t\) of the agent-based model variant (P=principal, A=agent).  }
  \label{fig:4}
\end{figure}

In the agent-based representation\footnote{We follow an approach introduced in \cite{guerrero2011} and \cite{Leitner2015a} to transfer closed-form mathematical models into agent-based models. In order to build a simulator from the model described in Sec. \ref{sec:abm} we use MathWorks\textsuperscript{\textregistered} Matlab.} of the hidden-action model we indicate time periods by subscript \(t = 1,2,...,T\) and the sequence of events within one timestep \(t\) by subscript \textcolor{black}{\(\tau=0,1,...,7 \)}. Figure \ref{fig:4} schematically represents the \textcolor{black}{sequence of events within one timestep in the }agent-based model and also indicates the \textcolor{black}{types of} information systems which the principal and the agent use during the different steps \(\tau\).

\paragraph{Characteristics of the principal, the agent, and the environment.} We characterize the (risk-neutral) principal by the utility function
%\
\textcolor{black}{
\begin{equation}
U_P\left(x_t - s (x_t )\right)=x_t-s(x_t)~,
\label{eq:1}
\end{equation}}

\noindent where \(x_t\) denotes the outcome and \textcolor{black}{\(s(x_t)=x_t\cdot p_t\), where \(p_t \in [0,1]\),} stands for the agent's compensation in \(t\). We model the agent as being  characterized by the productivity \(\rho\) \textcolor{black}{(which is constant throughout the simulations)}, we denote the agent's exerted effort in \(t\) by \(a_t\), and formalize outcome \(x_t\) in period \(t\) by 

\begin{equation}
\label{eq:prodfunction}
x_t=a_t\cdot \rho + \theta_t~.
\end{equation}

\noindent We model exogenous factors \(\theta_t\) follow a normal distribution, \(\theta_t\sim N(\mu,\sigma)\). As it is assumed in the standard hidden-action model (see \cite{holmstrom1979}), the principal aims at maximizing her utility subject to the participation constraint and the incentive compatibility constraint \textcolor{black}{(see Eqs. (\ref{eq1:maximization})-(\ref{eq1:ICC}))}: In order to do so, {we allow the principal to adapt the parameterization of the sharing rule, i.e. the premium parameter \(p_t\), over time}. 

The (risk-averse) agent is characterized by the CARA utility function 
\textcolor{black}{
\begin{equation}
U_A\left(s(x_t), a_t\right) = \frac{1-e^{-\eta \cdot s(x_t)}}{\eta} - 0.1{a_t^2}~,
\label{eq:2}
\end{equation}}

\noindent where \(\eta\) represents the agent's Arrow-Pratt measure of risk-aversion \cite{pratt1975}. 

\paragraph{Information about the action space \textcolor{black}{provided by} IS 2-P and IS 2-A.} In Sec. \ref{ha:is2} we introduce information asymmetry regarding the set of feasible actions. \textcolor{black}{In the agent-based model variant, we denote the set of feasible actions in timestep \(t\) by \(\mathbf{A}_t\). Limited information about \(\mathbf{A}_t\) hinders }the principal from recognizing the optimal \textcolor{black}{value for the premium parameter}, \(p_t\), \textcolor{black}{instantaneously}---this is in contrast to the standard \textcolor{black}{hidden-action} model, which suggests that the principal \textcolor{black}{always has the necessary information to set} the premium parameter \(p_t\) so that the agent has incentives to exert the optimal effort level (cf. Sec. \ref{sec:ha-nutshell} and Eq. (\ref{eq:premium-level}) below). As a consequence, the principal has to search for \textcolor{black}{the value of the premium parameter, which induces the optimal} effort level: In order to do so, the principal has the option to either exploit the \textcolor{black}{known} fraction of the action space (local search) or to explore the action space outside of the known area (global search) (cf. \cite{march1991}). In periods \(t=2,...,T\) the principal selects her search strategy in \(\tau=0\)  and, according to this strategy, uses her IS 2-P to perform either a global or a local search for effort levels on which she will base the \textcolor{black}{sharing rule \(s(\cdot)\)} (cf. \(\tau=0\) in Fig. \ref{fig:4}).\footnote{In \(t=1\), the principal randomly selects one candidate for the optimal effort level and bases the further procedure on this candidate.} 
In order to assure the existence of a solution to the principal's decision problem, it is necessary to set boundaries for \textcolor{black}{\(\mathbf{A}_t\)}. We \textcolor{black}{set} the lower boundary by the participation constraint, \textcolor{black}{\(E(U_A( s(x_t),a_t)\geq \underline{U}\)}, and the upper boundary by means of the incentive compatibility constraint, \textcolor{black}{\(a_t \in \argmax_{a'_t \in \bold{A_t}} E(U_A(s(x_t),a'_t))\) (cf. Eqs. (\ref{eq1:PC}) and (\ref{eq1:ICC}))}. It is important to note that both boundaries are endogenous: They include an expectation concerning the environmental variable, as \textcolor{black}{\(s(x_t)\)} is based on the outcome \textcolor{black}{\(x_t=a_t\cdot\rho + \theta_t\)} (cf. Eq. \ref{eq:prodfunction}). \textcolor{black}{The agent's reservation utility \(\underline{U}\), however, is an exogenous variable.} Changes in the state of information about the environment (via learning in \(\tau=7\), see below) might lead to changes in the boundaries of \textcolor{black}{\(\mathbf{A}_t\)}. The principal uses her IS 1-P to \textcolor{black}{compute her expectation} as to the environment. 

\paragraph{The principal's IS for external information (IS 1-P).} We denote the information which the principal retrieves from IS 1-P by the vector \textcolor{black}{\(\boldsymbol{\tilde{\Theta}}_t = [\tilde{\theta}_{t-1} \tilde{\theta}_{t-2} ... \tilde{\theta}_{t-m}  ]\). The elements of \(\boldsymbol{\tilde{\Theta}}_t\) stand for the principal's estimations of the exogenous factor in previous periods, for their computation see Eq. \ref{eq:estimation} below}. The length of  \(\boldsymbol{\tilde{\Theta}}_{t}\) is defined by parameter \textcolor{black}{\(m\in \{1,2,...,t \}\)}, which also stands for the sophistication of IS 1-P. A low level of sophistication indicates that only very recent information can be retrieved from IS 1-P, while a high level of sophistication means that information from a larger number of past periods is available for the principal.\footnote{Please note that the principal learns about the environment and stores the information acquired in her IS 1-P in \(\tau=7\) in each \(t\) (see Fig. \ref{fig:4}). This means that the principal only retrieves information from her IS that she entered into the system in previous periods \(r<t\), where \(0<r<t\), and also that the principal's information about the environment changes in each \(t\).} A higher value of \(m\), thus, indicates that the principal is better informed about the environment, as more historical information enables the principal to compute a more reliable expectation as to the environment.

\paragraph{Endogenous threshold to trigger the principal's search \textcolor{black}{strategy}.} As the information sources for the principal's decision for a search strategy are now defined (IS 1-P and IS 2-P), we can focus on her decision rule: The principal's decision to perform either a local or a global search (cf. \cite{march1991}) is based on (i) the principal's estimated exogenous factor in \(t-1\), \textcolor{black}{\(\tilde{\theta}_{t-1}\)}, and (ii) \textcolor{black}{her} propensity to innovate \(\delta\in[0,1]\).\footnote{For \(t=1\) the effort-level is a uniformly distributed random variable, see Fig. \ref{fig:4}.} \textcolor{black}{Recall that in periods \(t=2,3,...,T\) the} principal retrieves the (i) estimation of the exogenous factor from IS 1-P in \(\tau=0\). The (ii) propensity to innovate represents the principal's tendency to search either locally or globally for effort levels which can be used as the basis for the sharing rule \(s(\cdot)\). A lower (higher) value of \(\delta\) decreases (increases) the principal's tendency to search globally.  \textcolor{black}{Based on the principal's exploration propensity \(\delta\), we compute an exploration threshold \(\kappa_t\) for timestep \(t\) which is implicitly defined by }
\textcolor{black}{
 \begin{equation}
\delta = \dfrac{1}{\sigma(\boldsymbol{\tilde{\Theta}}_t)\cdot \sqrt{2\pi}} \cdot \int_{-\infty}^{\kappa_t} e^{-\frac{1}{2}\cdot\left(\dfrac{z-\mu(\boldsymbol{\tilde{\Theta}}_t)}{\sigma(\boldsymbol{\tilde{\Theta}}_t)}  \right)} dz ~,
\label{eq:3}
 \end{equation}}
 
\noindent where \(\sigma(\cdot)\) and \(\mu(\cdot)\) represent the standard deviation and the mean, respectively. If \(\tilde{\theta}_{t-1} > \kappa_t\) (\(\tilde{\theta}_{t-1} < \kappa_t\)) the principal performs a global (local) search.

\begin{figure}
 \includegraphics[width=0.9\linewidth]{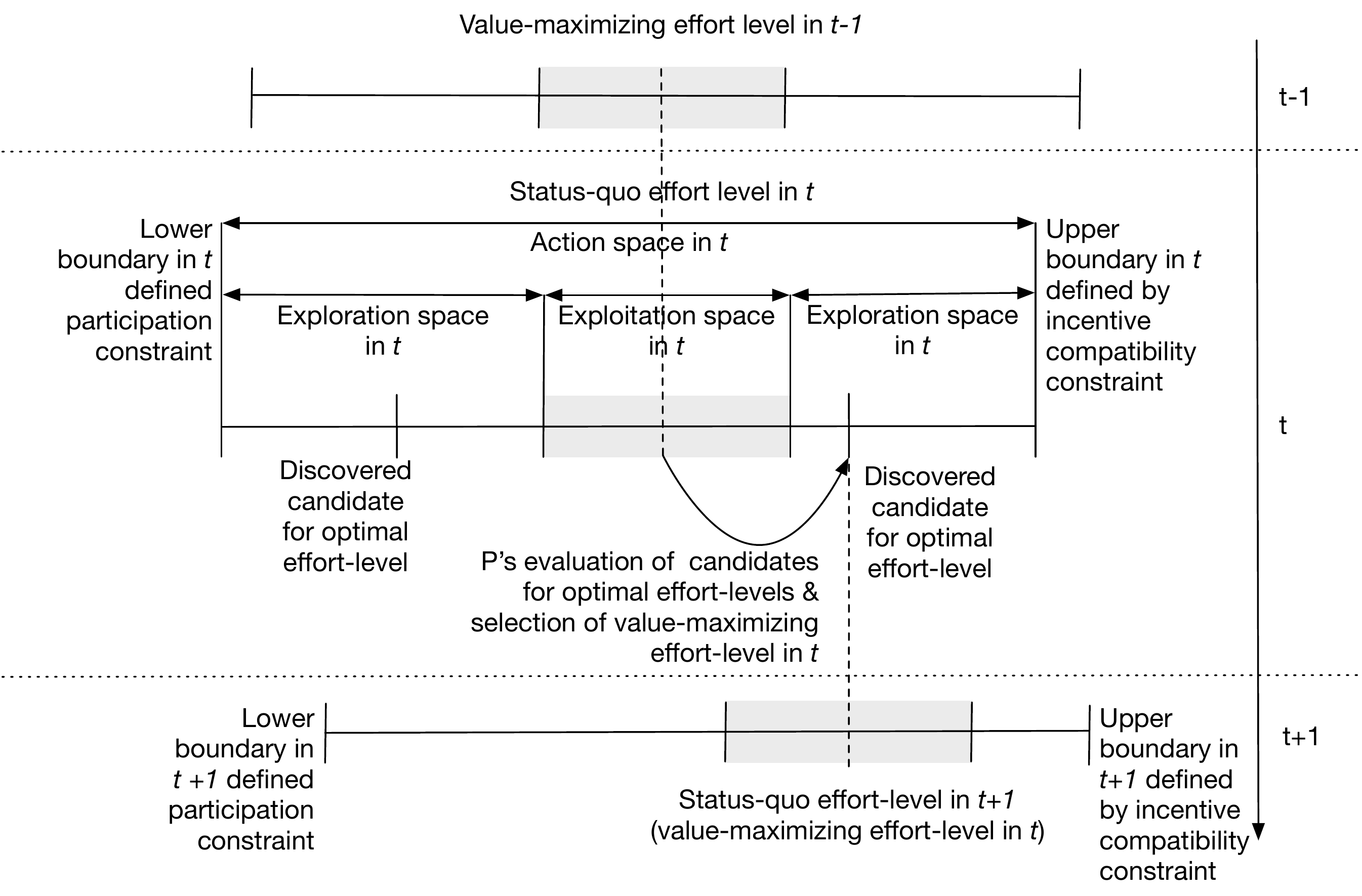}
  \caption{{Schematic representation of the endogenous boundaries of the action space (via the participation constraint and the incentive compatibility constraint), the definition of the principal's search spaces (exploration and exploitation space), and the interrelation between time-steps (indicated by dashed lines). The figure illustrates the case of exploration, in which the principal searches globally for candidates for the optimal effort level in \(t\).}}
 \label{fig:5}
\end{figure}

\paragraph{Endogenous boundaries of the principal's exploration and exploitation spaces.} The principal now has selected her search strategy for period \(t\). In order to carry out the search for candidates for the optimal effort level, the search spaces in which the principal performs her search for potential effort levels need to be defined. The search space for a local (global) search is referred to as exploitation (exploration) space (see also Fig. \ref{fig:5}). Please recall that the set of feasible actions \textcolor{black}{\(\mathbf{A}_t\)} is bounded by the incentive compatibility and the participation constraints. The search spaces can be defined as follows:
\begin{itemize}
\item  The \textbf{exploitation space} is defined as a fraction of the entire search space in \(t\) and is a consequence of the sophistication of P's IS 2-P. We denote the sophistication of IS 2-P by parameter \(q\), which defines the exploitation space in \(t\) as the fraction \(1/q\) of the entire action space \(\bold{A_t}\). We model the exploitation space to be equally distributed around the effort level on which the principal has based her computation of the premium parameter in the previous period, \(\tilde{a}_{t-1}\), and refer to \(\tilde{a}_{t-1}\) as the \enquote*{status-quo effort level}.
\item The \textbf{exploration space} is the area outside of the exploration space but inside the boundaries of \(\mathbf{A}_t\). 
\end{itemize}
The search spaces are schematically illustrated in Fig. \ref{fig:5}. Once the search strategy is \textcolor{black}{settled}, the principal randomly finds 2 alternative effort levels in the search space (\textcolor{black}{with uniformly distributed probabilities of discocery)}. The \textcolor{black}{discovered effort levels} will be evaluated in the next step. 

\paragraph{The principal's evaluation of effort levels.} The principal evaluates the newly found effort levels together with the status-quo effort level with respect to increases in her expected utility (based on the utility function in Eq. \ref{eq:1}) in timestep \(\tau=1\) (cf. Fig. \ref{fig:4}).\footnote{As first-order stochastic dominance is assumed for effort levels \cite{jost2001}, the principal will always opt for the highest discovered effort level.}  
We denote the value-maximizing effort level in \(t\) from the principal's point of view by \(\tilde{a}_t\).\footnote{In period \(t+1\) the effort level \(\tilde{a}_t\) will be referred to as status-quo effort level.} Please note that the fact that the principal evaluates all candidates for the optimal effort in \(t\) on the basis of Eq. \ref{eq:1}, requires the principal to build an expectation as to the environment. She retrieves this expectation from her IS 1-P in \(\tau=1\) (cf. Fig. \ref{fig:4}):
\textcolor{black}{
\begin{equation}
E_P(\theta_t) = \mu (\tilde{\Theta}_t )=
\begin{cases}
\frac{1}{t-1}~ \sum\limits_{n=1}^{n=t-1}\tilde{\theta}_{n} &\text{if~} m=\infty,\\
\frac{1}{m} \sum\limits_{\substack{\forall n \leq m: n=1 \\ \forall n > m: n = t-m }}^{n=t-1}\tilde{\theta}_{n} &\text{if~} m<\infty~.\\
\end{cases} 
\label{eq:theta}
\end{equation} 
}

\noindent Recall that parameter \(m\) indicates the sophistication of IS 1-P. The expected outcome \textcolor{black}{in timestep \(t\)} from the principal's point of view, \textcolor{black}{\(\tilde{x}_{Pt}\)}, using value-maximizing effort level \(\tilde{a}_t\) can, thus, be formalized by \(\tilde{x}_{t}=\tilde{a}_{t}\cdot\rho+E_P(\theta_t)\). 

\paragraph{The contract.} Now that the principal has decided on a desired effort level \(\tilde{a}_t\) for period \(t\), she can move on \textcolor{black}{fixing the incentive scheme \(s(\cdot)\)}. In order to do so, the principal computes the optimal premium-level in \(t\) in step \(\tau=2\) according to 
\textcolor{black}{
\begin{equation}
\label{eq:premium-level}
p_t=\max_{p\in[0,1]} U_P\left(\tilde{x}_{Pt},s(\tilde{x}_{Pt})\right)~.
\end{equation}}

\noindent She, then, designs the contract and offers the contract to the agent, who decides whether or not to accept it in \(\tau=3\). In order to compute the premium level \(p_t\), the principal uses information about the environment provided by IS 1-P and information about feasible actions provided by IS 2-P (cf. \(\tau=2\) in Fig. \ref{fig:4}). The agent uses IS 1-A and IS 2-A for his decision of whether to accept the contract or not (cf. \(\tau=3\) in Fig. \ref{fig:4}). 

\paragraph{The agent's IS for external information (IS 1-A) and his decision for an effort level.} \textcolor{black}{In case the agent accepts the contract, he exerts effort \textcolor{black}{\({a}_t = \max_{a_t \in \bold{A_t}} U_A\left(s(\tilde{x}_{At}), a_t\right)\)} in \(\tau=4\), where \(\tilde{x}_{At}=a_t \cdot \rho + E_A(\theta_t)\) includes the agent's expectation as to the outcome in \(t\), \(E_A(\theta_t)\).  In order to compute \(E_A(\theta_t)\), he retrieves his observations of realized exogenous factors from IS 1-A (cf. \(\tau=4\) in Fig. \ref{fig:4}). Recall that the agent can observe exogenous factors \textit{after} their realization. We denote the agent's observations derived from IS 1-A by {\(\boldsymbol{{\Theta}}_t = [{\theta}_{t-1} {\theta}_{t-2} ... {\theta}_{t-m}  ]\).} The agent computes his expectation as to the exogenous factor in \(t\) according to
\begin{equation}
E_A(\theta_t) = \mu(\boldsymbol{{\Theta}}_t)=
\begin{cases}
\frac{1}{t-1}~ \sum\limits_{n=1}^{n=t-1}{\theta}_{n} &\text{if~} m=\infty,\\
\frac{1}{m} \sum\limits_{\substack{\forall n \leq m: n=1 \\ \forall n > m: n = t-m }}^{n=t-1}{\theta}_{n} &\text{if~} m<\infty~.\\
\end{cases} 
\label{eq:theta-agent}
\end{equation} }

\noindent \textcolor{black}{As for IS 1-P, parameter \(m\) indicates the sophistication level of IS 1-A. The agent's \textcolor{black}{rule to come up with an effort level} introduced above also reflects that the agent \textcolor{black}{knows} the entire action space \(\mathbf{A}_t\). This piece of information is provided by IS 2-A (cf. \(\tau=4\) in Fig. \ref{fig:4}). }

\paragraph{\textcolor{black}{Realization of the outcome and of the principal's and the agent's utilities.}} After productive effort was exerted in \(\tau=4\) and the exogenous factor realized in \(\tau=5\),  the outcome, \(x_t = {a}_t \cdot \rho + \theta_t\), materializes in \(\tau=6\), and the utilities for the principal and the agent realize (cf. Eqs. \ref{eq:1} and \ref{eq:2}). 

\paragraph{The principal's estimation and agent's observation of the environment.} In \(\tau=7\), the agent observes the realization of the exogenous factor in \(t\), \(\theta_t\), and stores this observation in his IS 1-A. The principal can only observe \(x_t\) using IS 3. Based on this information she estimates the exogenous factor in \(t\) according to 
\begin{equation}
\tilde{\theta}_{t} = x_t - \tilde{a}_t \cdot \rho~, 
\label{eq:estimation}
\end{equation} 
and stores \(\tilde{\theta}_{t} \) in her IS 1-P.\footnote{Please note that, as long as only one piece of information is unavailable, \(\tilde{a}_t\) and \({a}_t\) perfectly coincide and, thus, the principal can estimate the realization of the exogenous factor without error.} Finally, \(\tilde{a}_t\) is carried over to period \(t+1\) as \enquote*{status-quo effort level}.\footnote{If the status-quo effort level is located outside the feasible region, the principal is forced to carry out a global search for alternative effort-levels.} This sequence explained above and depicted in Fig. \ref{fig:4} is repeated \(T\) times. \textcolor{black}{The notation used for the formal representation of the agent-based variant of the hidden-action model is summarized in Tab. \ref{tab:notation-abm}. }

 \begin{table}
 \textcolor{black}{
  \caption{Notation used in the (dynamic) agent-based model of the hidden-action problem}
   \label{tab:notation-abm}
 \begin{tabular}{p{0.57\textwidth}p{0.35\textwidth}}
\hline\noalign{\smallskip}				
Key element & Notation \\
\hline\noalign{\smallskip}
Timesteps																		&	\(t\)				\\
Index for sequence of events within one timestep										&	\(\tau\)			\\
Principal's utility function															&	\(U_P(x_t-s(x_t))=x_t-s(x_t)\)					\\
Principal's propensity to innovate													&	\(\delta\)			\\
Principal's exploration threshold in \(t\) 												&	\(\kappa_t\) 		\\
Agent's utility function															&	\(U_A(s(x_t),a_t)=V(s(x))-G(a)	\)				\\
Agent's utility from compensation													&	\(V(s(x_t))=\frac{1-e^{-\eta\cdot s(x_t)}}{\eta}\)		\\
Agent's Arrow-Pratt measure of risk-aversion											&	\(\eta\)			\\
Agent's disutility from effort														&	\(G(a_t)=0.1a^2_t\)	\\
Agent's reservation utility															&	\(\underline{U}\)		\\
Agent's productivity																&	\(\rho\)			\\
Agent's share of outcome	in \(t\)													&	\(s(x_t)=x_t\cdot p_t\) \\
Outcome in \(t\)																	&	\(x_t=a_t \cdot \rho + \theta_t \)					\\
Principal's expected outcome in \(t\)													&	\(\tilde{x}_{Pt} = \tilde{a}_t \cdot \rho + E_P(\theta_t)\) 	\\
Agent's expected outcome in \(t\)													&	\(\tilde{x}_{At} = {a}_t \cdot \rho + E_A(\theta_t)\) 	\\
Premium parameter in \(t\)															&	\(p_t\)			\\
Exerted effort level in \(t\)															&	\(a_t\)			\\
Value-maximizing effort level in \(t\) from the principal's perspective (and also status-quo effort level in \(t+1\))	&	\(\tilde{a}_t\)		\\
Set of feasible effort levels in \(t\)													&	\(\mathbf{A}_t\)		\\
Exogenous variable in \(t\)															&	\(\theta_t\)			\\
Principal's estimation of environmental variable in \(t\)									&	\(\tilde{\theta}_t\)	\\
Principal's information about the environment provided by IS 1-P in \(t\)						&	\(\tilde{\Theta}_t\) 	\\
Principal's expectation as to environmental variable in \(t\)								&	\(E_P(\theta_t)\)	\\
Agent's information about the environment provided by IS 1-A in \(t\)							&	\({\Theta}_t\) 		\\
Agent's expectation as to environmental variable in \(t\)									&	\(E_A(\theta_t)\)	\\
Sophistication parameter for IS 1-P													&	\(m\) 				\\
Sophistication parameter for IS 2-P													&	\(q\) 				\\
\hline\noalign{\smallskip} 
\end{tabular}
}
\end{table}

\subsection{Key parameters in the agent-based model}
\label{sec:sim}

\paragraph{Parameters related to the principal.} In addition to the utility function given in Eq. \ref{eq:1}, the principal is characterized by a propensity to innovate (\(\delta\)) which represents her tendency to perform a local or global search, respectively (see Eq. \ref{eq:3} and subsequent paragraphs). With respect to the propensity to innovate, our analysis includes three different types of principals:
\textcolor{black}{
\begin{enumerate}
\item  Exploitation-prone principals, who are characterized by a tendency towards local search (\(\delta=0.25\)). 
\item Indifferent principals, who assign equal probabilities to local and global search (\(\delta=0.5\)).
\item Exploration-prone principals, who are characterized by a tendency towards global search (\(\delta=0.75\)).
\end{enumerate}
}
\noindent \textcolor{black}{This parameter and the selected parameterization reflect current research on organizational ambidexterity that focuses on the ability of organizations to simultaneously pursue incremental and discontinuous innovation \cite{Tushman1996,Oreilly2013}. It is, amongst others, also noted in \cite{march1991} that organizations need to find an appropriate balance between exploration and exploitation to make sure to continuously adapt in order not to become irrelevant in the market. The principal's propensity for innovation reflects this balance between exploration and exploitation. }

\paragraph{Parameters related to the agent.} The agent is \textcolor{black}{risk-averse} and characterized by a CARA utility function (cf. Eq. \ref{eq:2}). \textcolor{black}{The assumption of risk-aversion for the agent is transferred from the standard hidden-action model \cite{holmstrom1979}.} We set the agent's Arrow-Pratt measure, \(\eta\), equal to \(0.5\). In addition, the agent is characterized by a measure for productivity, which we set to \(\rho=50\). 

\paragraph{Parameters related to the environment.} We model environmental variables to follow a normal distribution. In our analysis we set the mean of the distribution of environmental variables equal to zero and consider four levels of environmental turbulence, which we operationalize by altering the distribution's standard deviation,  \(\sigma\): We set the standard deviation relative to the optimal outcome, \(x^*\), computed by using actual parameter settings and the second-best solution of the standard hidden-action model \textcolor{black}{(cf. Appendix \ref{app:c})}, and consider a total of four cases ranging from of relatively stable environments (\(\sigma=0.05x^*\)) to relatively turbulent environments (\(\sigma=0.65x^*\)).\footnote{In order to compute the optimal outcome \(x^{*}\) we plug the principal's and the agent's utility functions (see Eqs. \ref{eq:1} and \ref{eq:2}, respectively) into the principal's optimization problem (see Eqs. \ref{eq1:maximization}-\ref{eq1:ICC}): We use the parameter settings of the actual scenario and compute the optimal sharing rule \(s(\cdot)\) and the corresponding outcome using MathWorks\textsuperscript{\textregistered} Matlab.}

\paragraph{Parameters related to ISs.} For the principal's and the agent's \textit{external information systems} (IS 1-P and IS 1-A, respectively), our analysis covers three levels of sophistication. Recall that the sophistication of these ISs is formalized by parameter \(m\) (see Eq. \ref{eq:theta} for IS 1-P and Eq. \ref{eq:theta-agent} for IS 1-A). We set \(m=1\) and \(m=3\) for a low and medium level of sophistication, respectively. For highly sophisticated ISs we set \(m=\infty\). The sophistication of P's \textit{internal information system} IS 2-P is captured by parameter \(q\), which identifies the fraction \(1/q\) of the set of feasible actions which is available for P. As discussed in Sec. \ref{ha:is2}, \(q\) is also a proxy for the extent of information asymmetry (regarding the action space) between the principal and the agent. Our analysis includes three sophistication levels of IS 2-P: We set \(q\in\{3,5,10\}\) for a high, medium, and low level of sophistication, respectively. \textcolor{black}{The choice of parameters reflects the concept of information introduced in Sec. \ref{ha:info}: Let \(J\) stand for the most complete information intrinsic to a system and \(I\) stand for information about a system and let observations about information intrinsic to a system be an information flow-process \(J \rightarrow I\) \cite{Frieden2010,Hawkins2010}. Then, the sophistication parameters \(m\) and \(q\) shape the extent of information (intrinsic to a system and observed or observable by the agent) which is available for the agent. Higher values for \(m\) (lower values for \(q\)) indicate that \textit{more} information is available, which we interpret as being better informed.}

\paragraph{Global parameters.} The possible combinations of the parameters which are subject to variation lead to a total number of \(3\times4\times3\times3 = 108\) scenarios. For each scenario \(R=700\) simulation runs are performed, whereby the analysis focuses on the first 20 time periods, \(T=20\).

\section{Results of the simulation study}
\label{sec:4}

\subsection{Effects of the level IS sophistication on the shape performance over time}
\label{sec:analysis1}

\begin{figure}
 \center \includegraphics[width=0.45\linewidth]{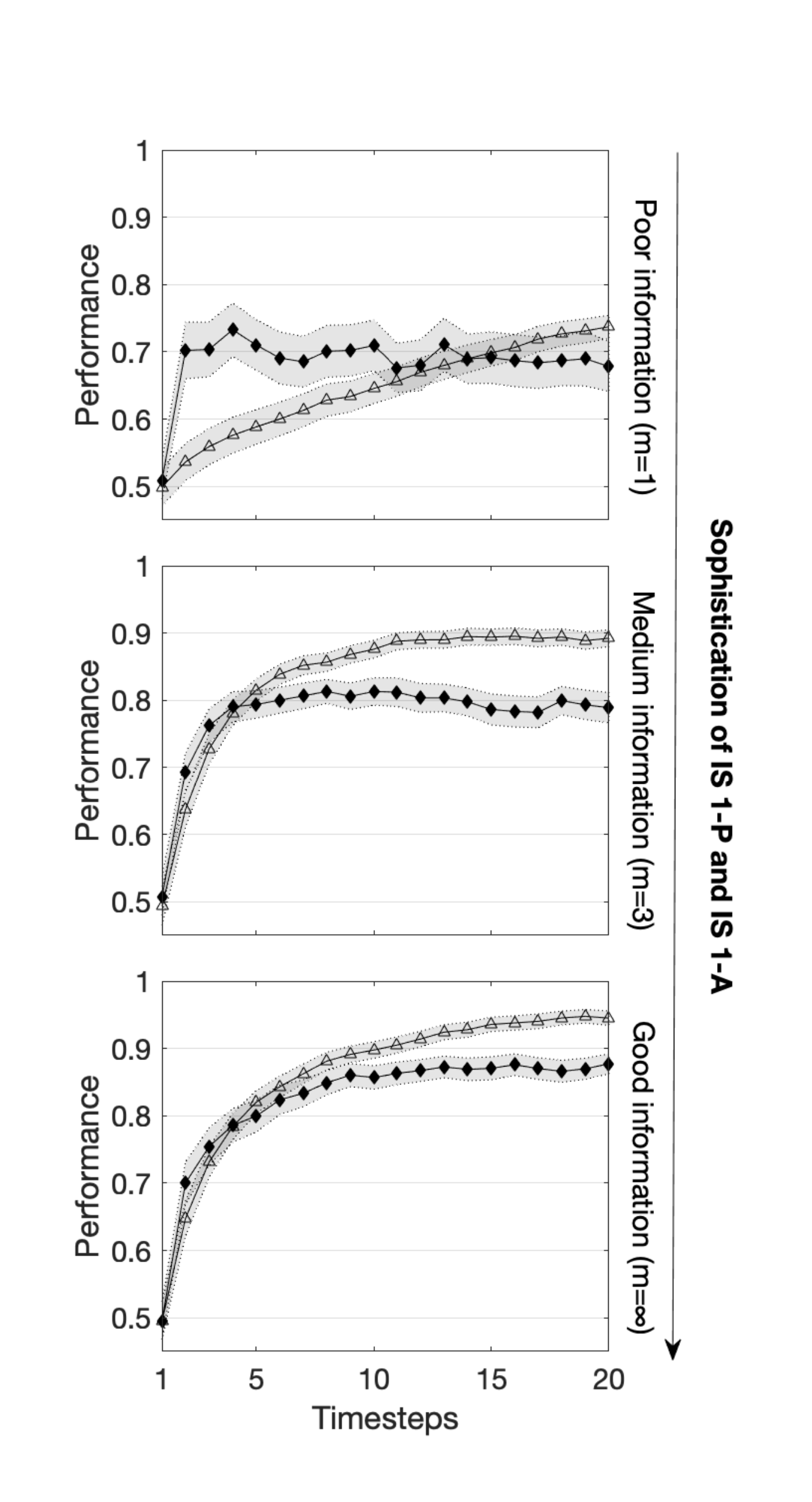}
  \caption{Effects of sophistication of the principal's and the agent's information systems for external information (IS 1-P and IS 1-A) on performance for different levels of environmental uncertainty. Averaged normalized effort levels are reported as performance measure (cf. Eq. \ref{eq:perf-measure}). The sophistication of the principal's information system for internal information (IS 2-P) is set to \(1/q=1/10\). Triangles (\(\triangle\)) represent low environmental uncertainty (\(\sigma=0.05x^*\)), black diamonds (\(\blacklozenge\)) represent high environmental uncertainty (\(\sigma=0.65x^*\)). Shaded areas indicate confidence intervals for \(\alpha=0.01\).}
  \label{fig:analyse-1b}
\end{figure}

\paragraph{Scenarios.} The first part of the analysis provides insights into the effects of the sophistication of the information systems employed within the contractual relationship between the principal and the agent on the level of performance obtained. In order to do so, this section presents results for the following parameter settings:
\begin{enumerate}
\item The sophistication levels \(m\) of the external information systems IS 1-P and IS 1-A are varied. We present results for 3 sophistication levels: ISs which provide poor (\(m=1\)), medium (\(m=3\)), and good information (\(m=\infty\)). Please note that \(m\) does not vary between the principal and the agent. 
\item The sophistication level \(1/q\) of the principal's internal information system IS 2-P is varied. The scenarios cover cases in which the IS 2-P provides poor (\(1/q=1/10\)), medium (\(1/q=1/5\)), and good information (\(1/q=1/3\)). 
\item The principal's propensity to innovate is set to a medium level, \(\delta=0.5\). This parameter setting results in the principal assigning equal probabilities to exploration and exploitation, while searching for candidates for the optimal effort level in \(t\).
\item We present results for the two extreme cases of low (\(\sigma = 0.05x^{*}\)) and high (\(\sigma = 0.65x^{*}\)) environmental uncertainty. 
\end{enumerate}
\noindent This section particularly discusses cases in which \(m\in\{1,3,\infty\}\) and \(1/q=1/10\) (cf. Fig. \ref{fig:analyse-1b}), and \(1/q\in\{1/10,1/5,1/3\}\) and \(m=1\) (cf. Fig. \ref{fig:analyse-1a}). The results for all parameter combinations are presented in Fig. \ref{fig:analyse-1} in Appendix \ref{app:a}.  

\paragraph{Performance indicator.} For each timestep \(t\), we report the averaged normalized effort level carried out by the agent as performance measure. For each timestep \textcolor{black}{\(t =1,...,T\)} and each simulation run \textcolor{black}{\(r =1,...,R\)} we track the level of effort \(a_{tr}\) exerted by the agent, and normalize it by the optimal level of effort \(a^*\). The optimal effort level results from the second-best solution suggested by the standard hidden-action model \textcolor{black}{(see \cite{holmstrom1979} and Appendix \ref{app:c})}. The reported performance indicator is formalized by 

\begin{equation}
\label{eq:perf-measure}
\tilde{p}_t = \frac{1}{R} \sum_{r=1}^{r=R} \frac{a_{tr}}{a^*}
\end{equation}

\paragraph{Results on the sophistication of IS 1-P and IS 1-A.}  The results on the sophistication of the principal's and the agent's systems for information about the environment are presented in Fig. \ref{fig:analyse-1b}. For this analysis, we investigate variations of IS 1-P and IS 1-A as described above, and keep the sophistication of the principal's IS 2-P constant at \(1/q=1/10\). Each subplot in Fig. \ref{fig:analyse-1b} presents results for one investigated sophistication level of IS 1-P and IS 1-A. For each scenario, the subplots report the averaged normalized effort level introduced in Eq. \ref{eq:perf-measure}. 
For scenarios with \textit{low environmental uncertainty} (represented by triangles in Fig. \ref{fig:analyse-1b}), the results indicate that increasing the sophistication of IS 1-P and IS 1-A---and thereby increasing information about the organization's environment---significantly increases the slope of the performance curves. Thus, effort for better information about the environment appears to pay off almost immediately in such scenarios. In addition to the slopes of the performance curves, the performances at the end of the observation period (i.e., after 20 periods) increase with the level of IS sophistication, too: While for the case of poor information about the environment around \(0.74\) of the performance suggested by the standard hidden-action model can be achieved after 20 periods (see the subtop plot in Fig. \ref{fig:analyse-1b}), increasing the sophistication so that good information is provided leads to a final performance of almost \(0.95\) (see the bottom subplot in Fig. \ref{fig:analyse-1b}). The described patterns can also be observed for situations in which the principal's internal IS 2-A has higher sophistication levels: However, the less it is pronounced the higher the sophistication of IS 2-P (see Fig. \ref{fig:analyse-1} in Appendix \ref{app:a}). 

As soon as we switch to scenarios with \textit{high environmental uncertainty} (represented by black diamonds in Fig. \ref{fig:analyse-1b}), similar shapes of the performance curves emerge: For the case of poor external information, for example, performance increases only in the first 2 periods to around \(0.68\), before it remains on this level until the end of the observation period.  In scenarios with good information about the environment, performance increases in the first 11 periods to around \(0.87\) and then remains stable until period 20. {From the results we can draw the conclusion that  higher sophistication levels of the IS for external information lead to (i) longer time spans in which performance increases, but (ii) the slopes of the performance curves in the first few periods are not affected. Consequently, final performances increase with the quality of the provided information.} The same pattern emerges for situations with higher sophistication levels of {P's IS 2-P} (cf. Fig. \ref{fig:analyse-1} in Appendix \ref{app:a})

\paragraph{Results on the sophistication of IS 2-P.} We depict the performance curves for variations in the sophistication level of the principal's IS for internal information IS 2-P in Fig. \ref{fig:analyse-1a}, whereby each of the 3 subplots presents results for one of the investigated sophistication levels. We keep the sophistication levels of IS 1-P and IS 1-A constant at \(m=1\).
 For scenarios in which \textit{environmental uncertainty is low} (\(\sigma=0.05x^*\), indicated by triangles in Fig.\ref{fig:analyse-1a}), our results suggest that increasing the sophistication of the principal's internal IS 2-P significantly increases the slope of the performance curve. Performance, thus, increases much faster in the first few periods. The final performance achieved after 20 periods is, however, only marginally affected: The final performance for the case of poor internal information (\(1/q=1/10\)) amounts to around \(0.71\) (see the right subplot in Fig. \ref{fig:analyse-1a}). Increasing the sophistication of IS 2-P to \(1/q=1/3\) only leads to a marginal increase, so that a final performance of around \(0.82\) can be achieved. For scenarios with a high sophistication of the principal's and the agent's information system for environmental information, the pattern is similar, but the increase in the slopes of the performance curves is less pronounced (cf Fig. \ref{fig:analyse-1} in Appendix \ref{app:a}).

A totally different pattern emerges in situations in which \textit{environmental uncertainty is high} (\(\sigma=0.65x^*\), indicated by black triangles in Fig. \ref{fig:analyse-1a}): Irrespective of the sophistication level of IS 2-P, performance \textit{immediately} reaches a level of around \(0.71\). From period \(t=2\) onwards, there are generally no significant changes on this performance level. A similar behavior can be observed for situations in which the principal and the agent are better informed about the environment (see Fig. \ref{fig:analyse-1} in Appendix \ref{app:a}). This is a remarkable and counter-intuitive result: The results suggest that---at least for the first few periods---a significantly higher level of performance can be achieved in turbulent environments, as compared to stable environments.This finding might be explained by the pressure to be innovative, which turbulent environments impose on the principal \cite{mendes2016}. The time span in which this result can be observed is critically shaped by the sophistication of the principal's internal information system: For \(1/q=1/10\) (\(1/q=1/3\)) the performance observed in stable environments exceeds the performance in turbulent environments after 10 (4) periods. 

\begin{figure}
 \center \includegraphics[width=1.1\linewidth]{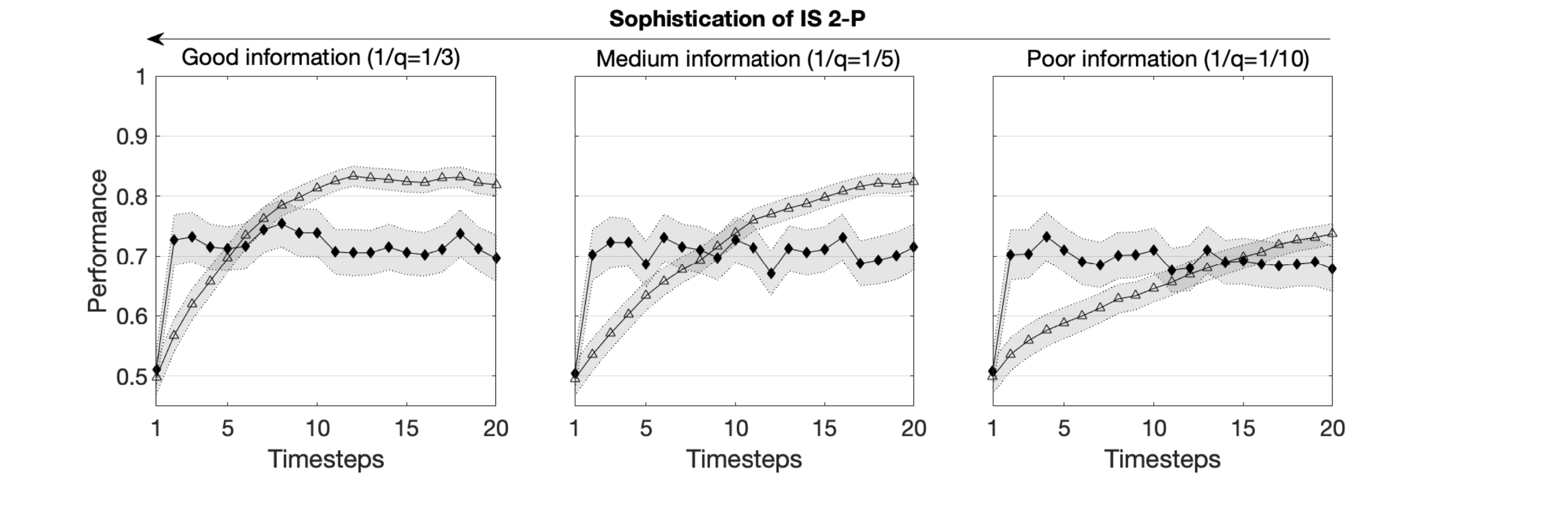}
  \caption{Effects of sophistication of the principal's information system for internal information (IS 2-P) on performance for different levels of environmental uncertainty. Averaged normalized effort levels are reported as performance measure (cf. Eq. \ref{eq:perf-measure}). The sophistication of the agent's and the principal's information systems for external information (IS 1-P and IS 1-A) is set to \(m=1\). Triangles (\(\triangle\)) represent low environmental uncertainty (\(\sigma=0.05x^*\)), black diamonds (\(\blacklozenge\)) represent high environmental uncertainty (\(\sigma=0.65x^*\)). Shaded areas indicate confidence intervals for \(\alpha=0.01\).}
  \label{fig:analyse-1a}
\end{figure}

\paragraph{Discussion and policy reflection.} The results provide support for intuition, i.e. that coping with environmental turbulence is more successful when information about the environment is improved \cite{raghun1999}. In this sense, the results are in line with prior research in the tradition of the task-technology model \cite{goodhue1995}: This line of research indicates that the relation between environmental uncertainty and task characteristics affects the satisfaction of the user with the data provided by ISs. This level of satisfaction, then, critically shapes the success of organizational ISs \cite{Karimi2004,Schaeffer2005,Petter2013}. In the model used in this paper, a higher sophistication of the ISs for environmental information (IS 1-P and IS 1-A) and the internal IS (IS 2-P) provides more complete information, which results in \enquote*{better} decisions.\footnote{Please note that this argumentation is only true as long as we consider decision makers who do not suffer from decision-making biases and are characterized by unlimited information-processing capabilities (cf., e.g., \cite{simon1957,tversky1974,raghun1999}).}  A higher sophistication of our modeled ISs can, thus, be interpreted as a higher task-technology fit. 

The results suggest that increasing the performance of the internal IS 2-P does not pay off in turbulent environment: This is an unexpected finding which puts calls for more sophisticated IS designs (e.g., \cite{Karimi2004}) into perspective. Our results suggest that turbulent environments put a certain pressure to be innovative on decision makers, which is why performance increases immediately in the first few periods. Further performance increases can, then, only be achieved by improving the quality of information about the environment (IS 1-P and IS 1-A), which is why efforts to increase the sophistication of the internal IS 2-P prove to be ineffective. 

The results presented so far suggest the necessity to differentiate the investment decisions regarding the sophistication of organizational ISs according to the degree of environmental uncertainty. If environmental uncertainty is low, enhancing the sophistication of external ISs (IS 1-P and IS 1-A) as well as internal IS 2-P increases the performance obtained: Investments in either direction appear to be beneficial. In contrast, when environmental uncertainty is high, investing into an improved internal IS 2-P appears to be ineffective: In such situations, the priority should rather be given to improving the quality of external information (provided by IS 1-P and IS 1-A). 

\subsection{Effectiveness of search strategies for different levels of IS sophistication}
\label{sec:analysis2}

\begin{figure}
 \center \includegraphics[width=0.45\linewidth]{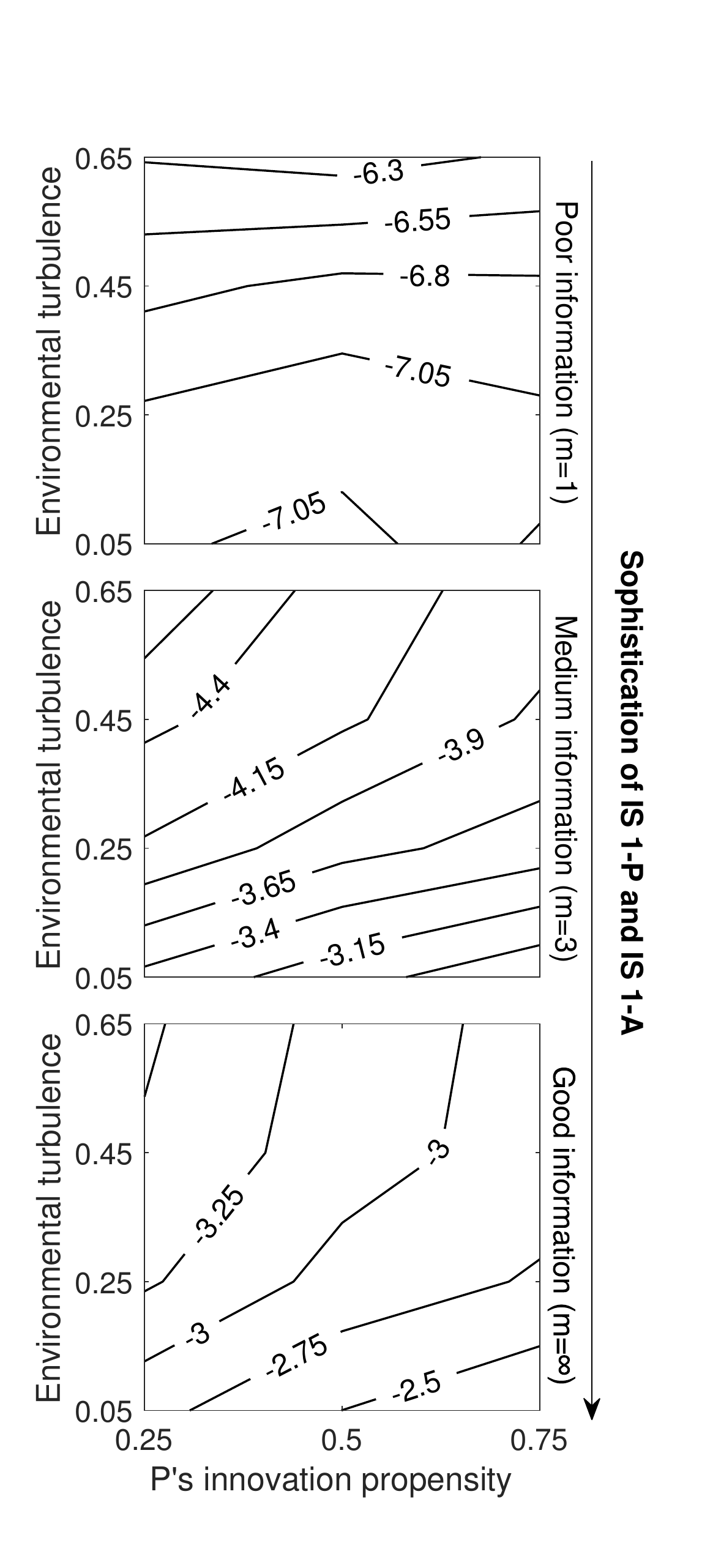}
   \caption{Effects of sophistication of the principal's and the agent's information systems for external information (IS 1-P and IS 1-A) on performance. Contours are based on Manhattan distances (cf. Eq. \ref{eq:perf-measure2}) between averaged normalized effort levels (introduced in Eq. \ref{eq:perf-measure}) and optimal performances \(x^*\). The sophistication of the principal's information system for internal information (IS 2-P) is set to \(1/q=1/10\).}
  \label{fig:analyse-2b}
\end{figure}

\begin{figure}[h]
 \center \includegraphics[width=1.1\linewidth]{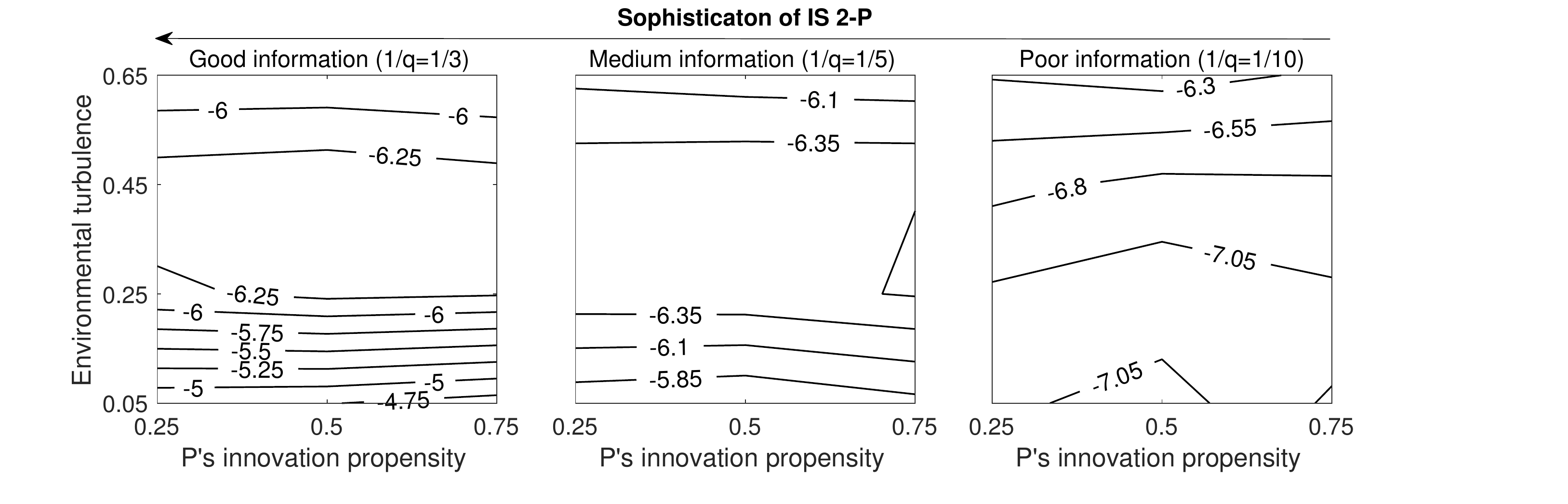}
    \caption{Effects of sophistication of the principal's information system for internal information (IS 2-P) on performance. Contours are based on Manhattan distances (cf. Eq. \ref{eq:perf-measure2}) between averaged normalized effort levels (introduced in Eq. \ref{eq:perf-measure}) and optimal performances \(x^*\). The sophistication of the principal's and the agent's information systems for external information (IS 1-P and IS 1-A) is set to \(m=1\).}
  \label{fig:analyse-2a}
\end{figure}

\paragraph{Scenarios.} This part of the analysis focuses not only on variations in the sophistication of the ISs employed during the principal's and the agent's decision-making processes, but also takes the principal's innovation propensity into account. The considered parameter setting is the following: \begin{enumerate}
\item As we did in Sec. \ref{sec:analysis1}, we vary the sophistication level \(m\) of the external information systems IS 1-P and IS 1-A, and analyze scenarios in which these ISs provide poor (\(m=1\)), medium (\(m=3\)), and good information (\(m=\infty\)).
\item Also the sophistication level (\(1/q\)) of the principal's internal information system IS 2-P is varied, so that the cases for poor (\(1/q=1/10\)), medium (\(1/q=1/5\)), and good information (\(1/q=1/3\)) are covered. 
\item We vary the principal's propensity to innovate: Like in Sec. \ref{sec:analysis1}, we investigate situations in which the principal is indifferent with respect to her search strategy (\(\delta=0.5\)). In addition, the results presented in this section cover situations in which the principal has a tendency for either exploration (\(\delta=0.75\)) or exploitation (\(\delta=0.25\)). 
\item We present results for 4 levels of environmental uncertainty: In addition to the two \enquote*{extreme} cases of low (\(\sigma = 0.05x^{*}\)) and high (\(\sigma = 0.65x^{*}\)) environmental uncertainty which were also included in the analysis in Sec. \ref{sec:analysis1}, we add two cases with intermediate environmental turbulence, in which we set  \(\sigma = 0.25x^{*}\) and \(\sigma = 0.45x^{*}\).
\end{enumerate}
\noindent The discussion in this section focuses on cases in which \(m\in\{1,3,\infty\}\) and \(1/q=1/10\) (cf. Fig. \ref{fig:analyse-2b}), and \(1/q\in\{1/10,1/5,1/3\}\) and \(m=1\) (cf. Fig. \ref{fig:analyse-2a}). The results for the remaining parameter combinations are presented in Fig. \ref{fig:analyse-2} in Appendix \ref{app:b}.  

\paragraph{Performance indicator.} The performance measure used for the analysis in this section is based on the level of effort exerted by the agent. We, however, no longer present the averaged normalized effort level \(\tilde{p}_{t}\) and the performance curves over time (as done in Figs. \ref{fig:analyse-1b} and \ref{fig:analyse-1a}), but condense performance curves to the Manhattan distance \(d\), which, in this case, represents the distance between the averaged normalized effort level \(\tilde{p}_t\) and the optimal effort level \(x^*\) over time. This allows us to present \textit{one} performance measure per scenario, which can be formalized by 

\begin{equation}
\label{eq:perf-measure2}
d=\sum_{t=1}^{t=T} \left(\tilde{p}_{1} -1 \right)~,
\end{equation}

\noindent where \(t=1,...,T\) represents time steps, and \(\tilde{p}_{t}\) stands for the averaged normalized effort level in period \(t\) (cf. Eq. \ref{eq:perf-measure}). 

\paragraph{Results on the sophistication of IS 1-P and IS 1-A.} Results for scenarios in which the sophistication levels of IS 1-P and IS 1-A for external information are varied are presented in Fig. \ref{fig:analyse-2b}. Each subplot presents the results for one of the investigated sophistication levels, and depicts contours resulting from the condensed performance measure introduced in Eq. \ref{eq:perf-measure2}. We keep the sophistication level of the principal's IS for internal information IS 2-P constant at \(m=1\) for this part of the analysis. 

As soon as we switch to scenarios with \textit{medium} (\(m=3\)) and \textit{high sophistication levels} (\(m=\infty\)) of the ISs for external information (IS 1-P and IS 1-A), we can observe that a different pattern emerges. First, the largest distance between the achieved and the optimal performances can no longer be observed in stable but in turbulent environments. This finding is in line with the intuition that there is a negative relation between environmental turbulence and performance. Second, we can see that the contours are no longer {nearly} horizontal but their slope increases with the quality of the information provided by the principal's and the agent's information systems for external information. This change in the pattern of contours indicates that the decision whether to perform exploitation or exploration becomes particularly critical, when the involved parties are well-informed about the environment. In addition to the changes in the pattern, it can be observed that the distance between the achieved and the optimal performance decreases significantly with increases in sophistication of the principal's Is 2-P: While for stable environments (\(\sigma=0.05x^*\)), indifferent principals (\(\delta=0.5\)), and poor external information (\(m=1\)) the Manhattan distance is around \(-7\), this distance reduces to around \(-3.15\) and \(-2.95\) for medium (\(m=3\)) and good (\(m=\infty\)) information, respectively. The marginal change in performance increase, thus, reduces with higher levels of IS sophistication. The same observations can be made for cases with a medium (\(1/q=1/5\)) and high (\(1/q=1/3\)) level of sophistication for IS 2-P (cf. Fig. \ref{fig:analyse-2} in \ref{app:b}). 

\paragraph{Results on the sophistication of IS 2-P.} Figure \ref{fig:analyse-2a} presents results from scenarios with variations in the sophistication level of the principal's internal IS 2-P. For the ISs for external information, we fix \(m=1\) for this part of the analysis. 
As discussed above, for the scenarios with a \textit{low sophistication level} of IS 2-P (\(1/q=1/10\)), the largest distance between the achieved performances and the optimal performance can be observed in stable environments, and the horizontal contours indicate that the choice of strategy does not affect performance.\footnote{Recall, if the sophistication of the information systems for external information increases to \(m=3\) and higher, this pattern can no longer be observed.} An increase in the sophistication level of the principal's IS for external information so that \textit{medium} (\(1/q=1/5\)) and \textit{good information} (\(1/q=1/3\)) is provided for decision-making purposes, only leads to slight changes in the observed pattern: First, the largest values for the distance from the achieved to the optimal performance shifts into the direction of more turbulent environments. The largest distances can, however, be observed for intermediate levels of environmental turbulence. This is surprising as one would expect the largest Manhattan distances in scenarios with the highest level of environmental turbulence. Second, for all sophistication levels of IS 2-P we can observe nearly horizontal contours: This indicates that, irrespective of the quality of internal information, the principal's search strategy does \textit{not} affect performance, as long as the sophistication of the IS for external information is \(m=1\). For higher sophistication levels of the ISs for external information (IS 1-P and IS 1-A) it can, however, be observed that a higher tendency for exploration leads to slightly better performances (cf. Fig. \ref{fig:analyse-2} in Appendix \ref{app:b}). 

\paragraph{Discussion and policy reflection.}  This section does not take a snapshot of one time period or analyze performance curves over time but provides a condensed measure for the efficiency of organizational search strategies and sophistication levels of information systems for time periods. From that perspective, the results presented here indicate that the intuition that it is harder for organizations to achieve a high performance in turbulent environments is only true if decision makers are well informed about the environment (i.e, in situations in which IS 1-P and IS 1-A provide medium or good information). In situations in which the principal and the agent only have poor information about the environment, there appears to be a pressure to be innovative in terms of carrying out tasks. This leads to an immediate boost in performance, so that---over the entire observation period---the distance between the achieved and the optimal performance decreases. This finding is in line with previous research: Eisenhardt \cite{eisenhardt1989b} and Alexiev et al. \cite{alexiev2016}, for example, argue that increased innovativeness is a common response of organizations to turbulent environments when information quality is poor. This is exactly what we  observe for situations in which the principal and the agent have only limited information about the environment. In addition, our results show that this pressure does not exist in stable environments, which is why performance increases at a much slower pace. In addition, we analyze the effect of the search strategy's impact on performance: Auh and Menguc  \cite{auh2005} argue that whether or not exploration or exploitation is the superior search strategy depends on the type of organization, which, in their case, is either defender or prospector. We show that as long as information about the environment is poor, the choice of the search strategy in fact does not matter. With an increase in quality of information about the environment, exploration becomes significantly superior to exploitation. For the sophistication of the principal's information system for information about the action space, the results indicate that investments into highly sophisticated ISs only lead to very marginal increases of performance and no changes in the above discussed patterns. Thus, from a policy perspective, the findings presented here suggest a prioritization of ways to spend an organization's resources: Every effort should be made to build an information system which provides good information about the environment before tackling the question of whether to develop new ways to carry out specific tasks. 

\section{Summary and conclusive remarks}
\label{sec:5}

The standard hidden-action model (see \cite{holmstrom1979}) comprises some rather \enquote*{heroic} assumptions about the availability of information and individual behavior. In this paper, we put a particular focus on the assumptions regarding the information which is accessible for both the principal and the agent. We relax selected assumptions and, by doing so, shift the focus from the decision-influencing role to the decision-facilitating role of information. For this purpose, we employ an approach allowing the transfer of closed-form mathematical models into agent-based models \cite{guerrero2011,Leitner2015a}, which allows to make less restrictive assumptions.

We limit the principal's and the agent's information about the environment in which the organization operates but endow them with the ability to learn about the environment over time. Both the principal and the agent store their information acquired in {an information system}. In addition, we add information asymmetry regarding the options available for the agent to carry out the task which the principal delegates to him: The principal is no longer fully informed about all feasible options but is endowed with search strategies to discover new options \cite{march1991}. This information asymmetry between the principal and the agent is operationalized by granting them access to two different information systems for this type of information. We also investigate different levels of information system sophistication and analyze the impact on performance. 

The results of our simulation study generate some key insights into the dynamics of delegation relationships with hidden-action:
\begin{itemize}
\item First, our results provide evidence of the intuition that coping with environmental turbulence is more successful when the quality of information about the nature of the environment is improved. We also observe that turbulent environments appear to put a pressure to be innovative on decision makers, which results in almost \enquote*{immediate} performance increases. Marginal changes in performance, however, decrease very fast, so that no further performance increases can be observed after only very few periods. 
\item Second, the results indicate that increasing the quality of internal information about feasible ways to carry out tasks significantly affects the levels of achieved performances.
\item Third, the results show that the choice of organizational search strategy (exploration or exploitation) affects performances only if decision makers are well informed about the environment. For the case of a poor quality of information about the environment, the employed search strategy does not significantly affect the level of achieved performance. 
\end{itemize}

Our research is, of course, not without its limitations. First, we take over some assumptions regarding the principal and the agent from principal-agent theory. These assumptions cover, for example, the individual utility-maximizing behavior or the availability of information about the agent's characteristics. Future research might address these assumptions and assess their impact on the applicability of incentive mechanisms provided by principal-agent theory. Second, the agent is modeled to carry out the same task repeatedly; we, however, assume that there are no learning-curve effects. Making the agent's productivity an endogenous variable might add further dynamics to the model. Third, we model situations in which no search costs occur for the exploration of the search space. Future research might also consider adding search costs. In addition, coming up with alternative incentive schemes which a promising line for future research.

 \section*{Conflict of interest}
The authors declare that they have no conflict of interest.

\appendix
\newpage
\section{IS sophistication and performance over time}
\label{app:a}

\begin{figure}[H]
 \center \includegraphics[width=\linewidth]{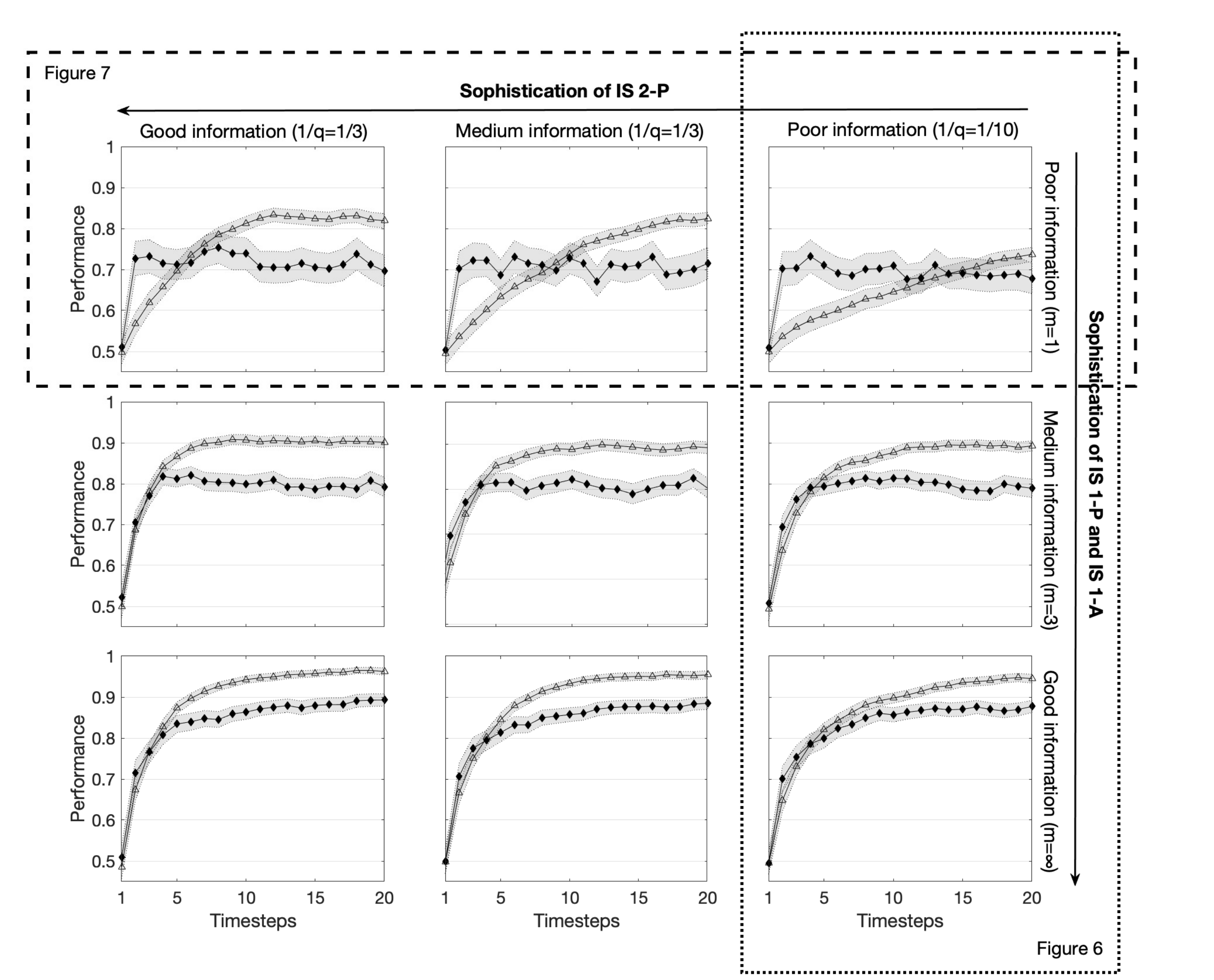}
  \caption{Effects of sophistication on performance for different levels of environmental uncertainty. Averaged normalized effort levels are reported as performance measure (cf. Eq. \ref{eq:perf-measure}). Triangles (\(\triangle\)) represent low environmental uncertainty (\(\sigma=0.05x^*\)), black diamonds (\(\blacklozenge\)) represent high environmental uncertainty (\(\sigma=0.65x^*\)). Shaded areas indicate confidence intervals for \(\alpha=0.01\). The dotted and dashed lines indicate the parts of the figure already presented in Figs. \ref{fig:analyse-1b} and \ref{fig:analyse-1a}, respectively.}
  \label{fig:analyse-1}
\end{figure}

\newpage
\section{IS sophistication and the effectiveness of search strategies}
\label{app:b}

\begin{figure}[H]
 \center \includegraphics[width=\linewidth]{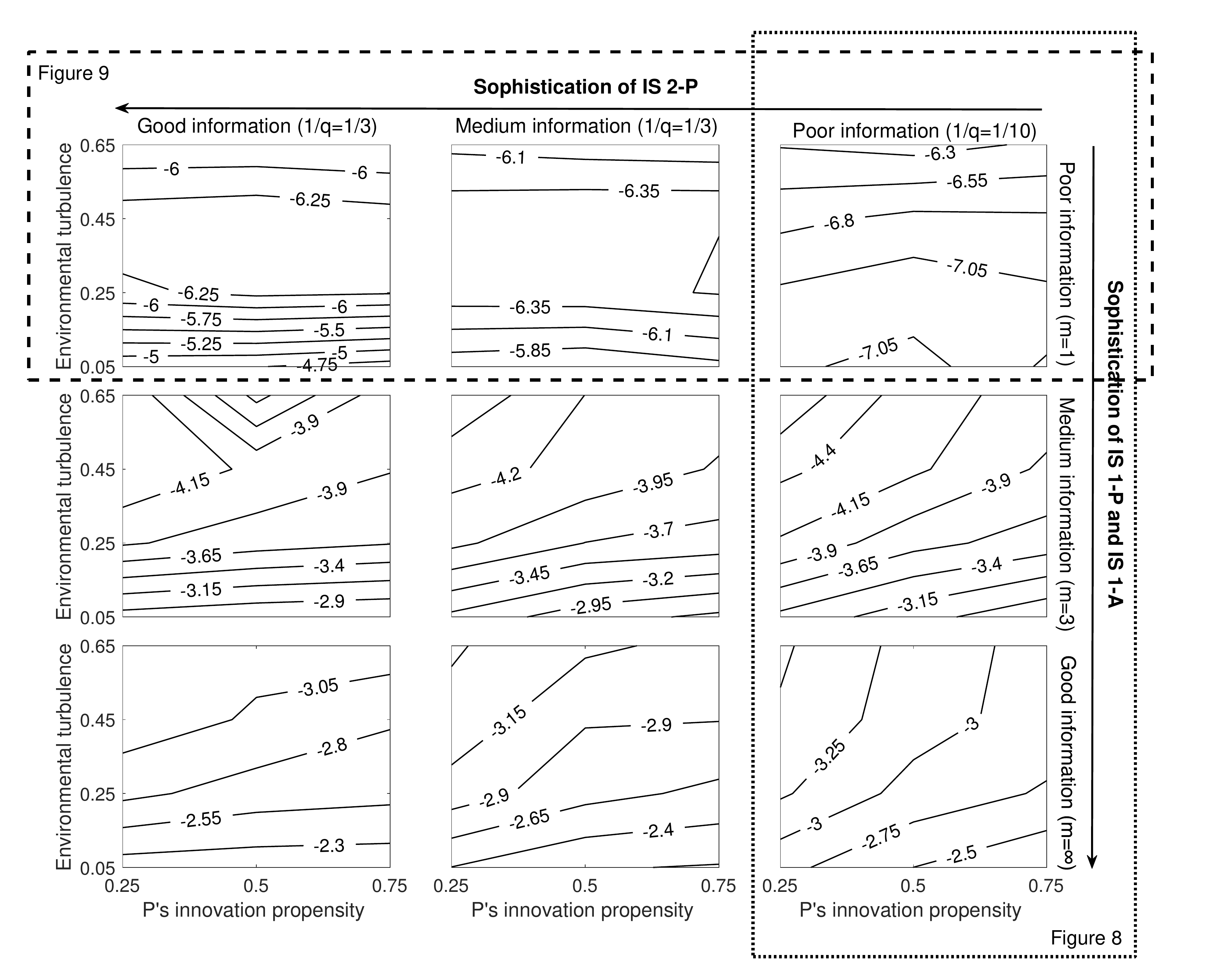}
    \caption{Effect of the principal's search strategies on performances for different levels of sophistication of information systems for internal and external information. Contours are based on Manhattan distances (cf. Eq. \ref{eq:perf-measure2}) between averaged normalized effort levels (introduced in Eq. \ref{eq:perf-measure}) and optimal performances \(x^*\). The dotted and dashed lines indicate the parts of the figure already presented in Figs. \ref{fig:analyse-2b} and \ref{fig:analyse-2a}, respectively.}
  \label{fig:analyse-2}
\end{figure}
\newpage

\section{Solution to the standard hidden-action model}
\label{app:c}
\textcolor{black}{For the solution to the optimization program introduced in Eqs. (\ref{eq1:maximization}) - (\ref{eq1:ICC}), Holmstr\"om \cite{holmstrom1979} suppresses the random state of nature \(\theta\) and views outcome \(x\) as random variable with distribution \(F(x,a)\), that is parameterized by the agent's effort \(a\) (see also \cite{Mirrlees1974,Mirrlees1976}). Given a distribution of the random state of nature, \(F(x,a)\) is this distribution induced on the outcome \(x\) via the function \(x=x(a,\theta)\). Holmstr\"om \cite{holmstrom1979} further assumes that a change in the effort level has an effect on the distribution of the outcome, \(F_a(x,a)>0\). Moreover, \(F(x,a)\) has a density function where \(f_a(x,a)\) and \(f_{aa}(x,a)\) are well defined for all \((x,a)\). Following the first-order approach (see, for example, \cite{Rogerson1985}), the optimization program introduced in Eqs. (\ref{eq1:maximization}) - (\ref{eq1:ICC}) can be reformulated as
\begin{subequations}
\begin{align}
\max_{s(x),a} 	\quad 	& 	\int U_P\left(x-s\left(x\right)\right)f(x,a)dx \label{appEq:maximization}\\
\textrm{s.t.} 			\quad 	&	\int [V(s(x)) - G(a)]f(x,a)dx \geq\underline{U} \label{appEq:PC}\\
 					 		&	\int V(s(x))f_a(x,a)dx = G'(a) \label{appEq:ICC}
\end{align}
\end{subequations}
We denote the multipliers for Eqs. (\ref{appEq:PC}) and (\ref{appEq:ICC}) by \(\lambda\) and \(\mu\), respectively. Point-wise optimization leads to the following characterization for the optimal sharing rule:
\begin{equation}
\label{eq:app-solutiona} 
\frac{U_P^\prime(x-s(x))}{V^\prime(s(x))}=\lambda + \mu \cdot \frac{f_a(x,a)}{f(x,a)}~.
\end{equation} 
For risk-neutral principals, Eq. (\ref{eq:app-solutiona}) reduces to 
\begin{equation}
\label{eq:app-solutionb} 
\frac{1}{V^\prime(s(x))}=\lambda + \mu \cdot \frac{f_a(x,a)}{f(x,a)}~.
\end{equation} 
In situations in which a first-best solution can be achieved, i.e. in situations in which the principal can observe the agent's effort, no incentive problem exists, which is why the incentive compatibility constraint given in Eq. (\ref{appEq:ICC}) is not binding and \(\mu=0\) in Eqs. (\ref{eq:app-solutiona}) and (\ref{eq:app-solutionb}). In such situations a fixed compensation for the agent is optimal. If there is an incentive problem and the principal wants the agent to increase his effort, Eq. (\ref{appEq:ICC}) must be binding and \(\mu>0\) in  Eqs. (\ref{eq:app-solutiona}) and (\ref{eq:app-solutionb}).} 

\textcolor{black}{
For further details, the reader is referred to \cite{holmstrom1979}. A further discussion of the standard hidden-action model and its solution is, for example, provided in \cite{Lambert2001} and \cite{Caillaud2000}.
}

\bibliographystyle{spphys}       % APS-like style for physics
\bibliography{bibliography}   % name your BibTeX data base

\end{document}